\documentclass[twocolumn,times, trackchanges]{aastex701}

\usepackage{multirow}
\usepackage{amssymb,amsmath}
\shorttitle{The Silk Nebula: A Newly Discovered Giant [\ion{O}{3}] Emission Filament }
\shortauthors{Yi-Fei Wu, Song-Yu Jia et al.}
\begin{document}

\title{The Silk Nebula: A Newly Discovered Giant [\ion{O}{3}] Emission Filament}

\author[0009-0000-9486-9548]{Yi-Fei Wu}
\affiliation{National Astronomical Observatories, Chinese Academy of Sciences, Beijing 100101, People’s Republic of China}
\affiliation{School of Astronomy and Space Sciences, University of Chinese Academy of Sciences, Beijing 100049, China}
\email{1964668672@qq.com}

\author[0009-0006-5531-536X]{Song-Yu Jia}
\affiliation{AEROSPACE KNOWLEDGE PRESS}
\email{jiasongyu163@gmail.com}

\correspondingauthor{Chao-Jian Wu}
\author[0000-0003-3514-6619]{Chao-Jian Wu}
\affiliation{National Astronomical Observatories, Chinese Academy of Sciences, Beijing 100101, People’s Republic of China}
\email[show]{chjwu@bao.ac.cn}

\correspondingauthor{Chao Zhang}
\author{Chao Zhang}
\affiliation{National Astronomical Observatories, Chinese Academy of Sciences, Beijing 100101, People’s Republic of China}
\email[show]{chaozhang@bao.ac.cn}

\author{Li-Yue Zhang}
\affiliation{National Astronomical Observatories, Chinese Academy of Sciences, Beijing 100101, People’s Republic of China}
\affiliation{School of Astronomy and Space Sciences, University of Chinese Academy of Sciences, Beijing 100049, China}
\email{lyzhang@bao.ac.cn}

\author{Shu Wang}
\affiliation{National Astronomical Observatories, Chinese Academy of Sciences, Beijing 100101, People’s Republic of China}
\email{shuwang@nao.cas.cn}

\author{Juan-Juan Ren}
\affiliation{National Astronomical Observatories, Chinese Academy of Sciences, Beijing 100101, People’s Republic of China}
\email{jjren@nao.cas.cn}

\author[0009-0000-7365-074X]{Quzhen Dongyang}
\affiliation{School of Astronomy and Space Science, Nanjing University, Nanjing 210093, People’s Republic of China}
\affiliation{Key Laboratory of Modern Astronomy and Astrophysics, Nanjing University, Ministry of Education, Nanjing 210093, People's Republic of China}
\email{ChoedronD@proton.me}

\author{Hong-Yun Qiu}
\affiliation{Light Speed Vision Beijing Co.LTD (QHYCCD)}
\email{1964668672@qq.com}

\author{Zhi-Hong He}
\affiliation{School of Physics and Astronomy, China West Normal University, Nanchong, China}
\email{hezh@cwnu.edu.cn}

\author[0000-0002-0971-9535]{Jichen Zhang}
\affiliation{Department of Physics, The Chinese University of Hong Kong, Shatin, N.T., Hong Kong, China}
\email{jczhang@link.cuhk.edu.hk}

\author{Guo-You Sun}
\affiliation{Xingming Observatory, Urumqi, Xinjiang, People's Republic of China}
\email{1964668672@qq.com}

\author{Wei-Guo Zhang}
\affiliation{Lijiang Gemini Observatory}
\email{1964668672@qq.com}

\author{Jian-Jun Chen}
\affiliation{National Astronomical Observatories, Chinese Academy of Sciences, Beijing 100101, People’s Republic of China}
\email{jjchen@nao.cas.cn}

\author[0000-0002-8609-3599]{Hong-Liang Yan}
\affiliation{National Astronomical Observatories, Chinese Academy of Sciences, Beijing 100101, People’s Republic of China}
\affiliation{School of Astronomy and Space Sciences, University of Chinese Academy of Sciences, Beijing 100049, China}
\email{hlyan@nao.cas.cn}

\author{Xiang-Song Fang}
\affiliation{National Astronomical Observatories, Chinese Academy of Sciences, Beijing 100101, People’s Republic of China}
% \affiliation{CAS Key Laboratory of Optical Astronomy, National Astronomical Observatories, Chinese Academy of Sciences, Beijing 100101, People's Republic of China}
\email{xsfang@bao.ac.cn}

\author{Hong Wu}

% \affiliation{CAS Key Laboratory of Optical Astronomy, National Astronomical Observatories, Chinese Academy of Sciences, Beijing 100101, People's Republic of China}
\affiliation{National Astronomical Observatories, Chinese Academy of Sciences, Beijing 100101, People’s Republic of China}
\affiliation{School of Astronomy and Space Sciences, University of Chinese Academy of Sciences, Beijing 100049, China}
\email{hwu@bao.ac.cn}
\collaboration{all}{}

\begin{abstract}

\noindent We report [O\,\textsc{iii}] narrow-band imaging of an extremely large-scale nebula, the \textit{Silk Nebula}, discovered by the SWONS (Super Wide-Field [O\,\textsc{iii}] Narrow-band Survey), together with multi-fiber medium-resolution spectroscopic data covering the H$\alpha$, [N\,\textsc{ii}], [S\,\textsc{ii}], and [O\,\textsc{iii}] emission lines, acquired with the Large Sky Area Multi-Object Fiber Spectroscopic Telescope (LAMOST) as part of the Medium-Resolution Spectroscopic Survey of Nebulae (MRS-N). 
Based on 2,684 high-quality spectra, we measured its physical parameters, including relative line intensities, radial velocities (RVs), and full widths at half-maximum (FWHMs).
The [S\,\textsc{ii}]/H$\alpha$ ratio of the nebula significantly exceeds the shock-excitation threshold of 0.4. Its location in BPT diagnostic diagrams and spatial morphology of [O\,\textsc{iii}] both support the optical emission is dominated by shock-heated gas. The oxygen abundance is near solar and spatially uniform, consistent with a shock propagating into the ambient interstellar medium. Because [O\,\textsc{iii}] is the only emission line that clearly traces the nebular morphology, we base our distance estimate exclusively on the [O\,\textsc{iii}] kinematics, obtaining a fiducial distance of $\sim$160 pc. At this distance, the nebula has a physical extent of $\sim$28 pc along the north-south direction. We suggest that the Silk Nebula is a large [O\,\textsc{iii}] filament possibly associated with a supernova remnant, although a diffuse-ionized-gas (DIG) or ionized-bubble origin cannot be ruled out with the present data, and one of the largest angular-scale [O\,\textsc{iii}] emission structures documented to date.
% However, the classification is complicated by elevated [S\,\textsc{ii}]/H$\alpha$ ratios in the background region and a systematic velocity offset between the [O\,\textsc{iii}] and low-ionization lines, indicating significant DIG contamination along the line of sight. Taken together, we identify the \textit{Silk Nebula} as a candidate supernova remnant, one of the largest angular-scale [O\,\textsc{iii}] structures documented to date. 
Future deep X-ray observations and shock-modeling efforts are required to firmly establish its nature.

\end{abstract}
\keywords{\uat{Supernova remnants}{1667} --- \uat{Emission nebulae}{461} --- \uat{Interstellar medium}{847} --- \uat{Spectroscopy}{1558}}


\section{Introduction} 
\label{sect:intro}
%---------------------------------------------------------------
\begin{figure*}[h!]
   \centering
   \includegraphics[width=1\textwidth]{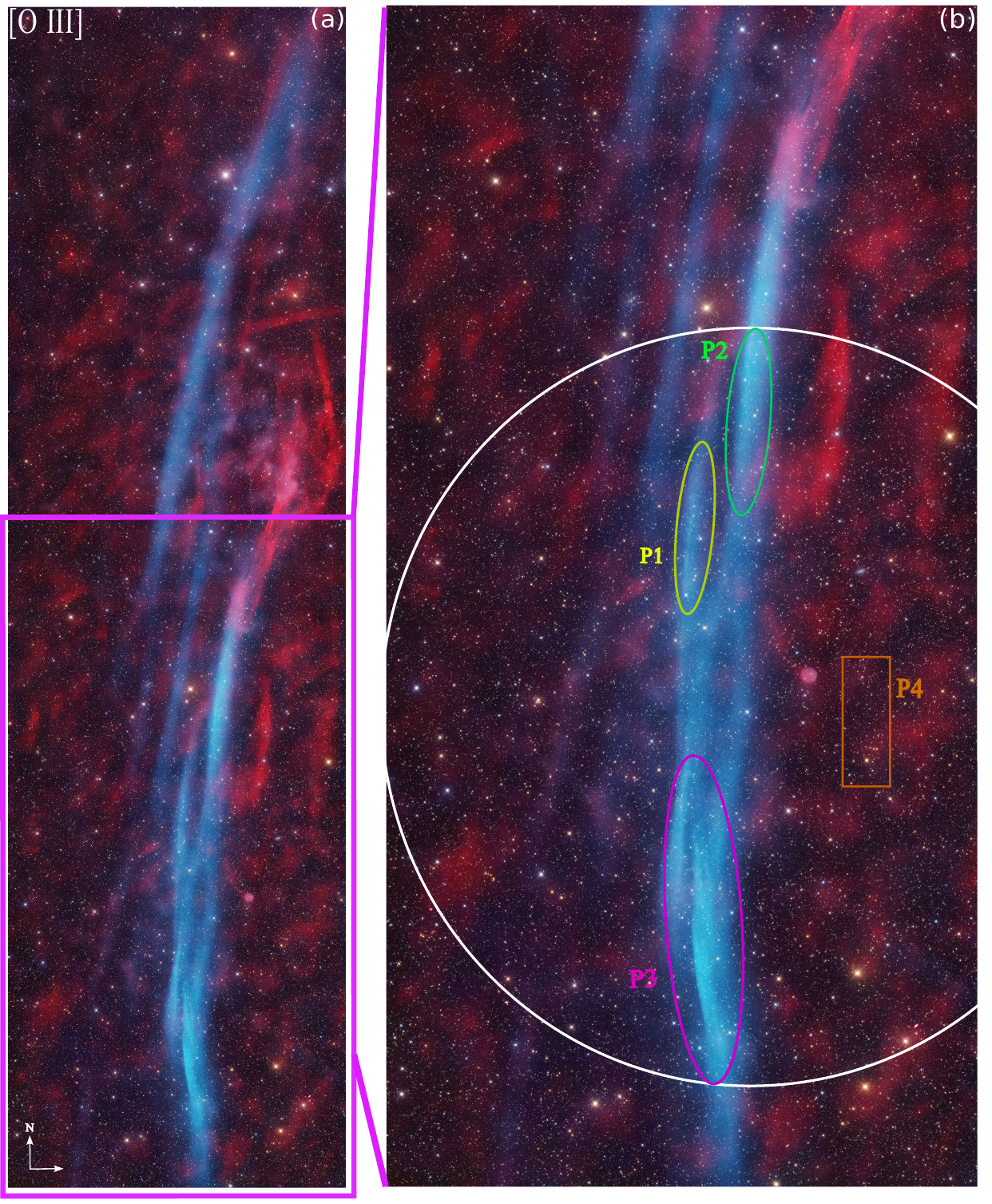}
    % \caption{\textcolor{red}{Deep color composite images of Silk Nebula, color composite of H$\alpha$ (red) and [O\,\textsc{iii}] (blue): (a) The entire nebula image obtained by SWONS, with the image FOV of 1$^\circ$ $\times$ 10$^\circ$. (b) Blowup view of the region near the LAMOST field of view (white circle), dividing into three regions (P1-P3) based on [O\,\textsc{iii}] intensity distribution, with region P4 (no detected [O\,\textsc{iii}] emission) shown for comparison as background. }}
    \caption{Deep color composite images of the \textit{Silk Nebula}, combining H$\alpha$ (red) and [O\,\textsc{iii}] (blue) emission. (a) The full nebula image obtained by the SWONS team, with a field of view of 1$^\circ$ $\times$ 10$^\circ$. (b) Enlarged view of the region near the LAMOST MRS-N field of view, which is indicated by the white circle. We divide this area into four regions for spectral analysis: P1-P3 are zones of relatively bright [O\,\textsc{iii}] emission, while P4 is a nebula-free region used as a background reference.}
   \label{fig:OIIIMAP.png}
\end{figure*}
%---------------------------------------------------------------
Ionized nebulae, such as Supernova remnants (SNRs), planetary nebulae (PNe), H\,\textsc{ii} regions and diffuse ionized gas (DIG), constitutes the principal components of emission nebulae and serves as key probes for elucidating stellar evolution, energy transfer processes, and the chemical evolution of the interstellar medium (ISM).
Over the past decades, numerous studies have investigated their physical mechanisms \citep{osterbrock2006astrophysics,vink2020physics,jin2022theoretically,fesen2023spatial,slane2013new,sharpless1953catalogue,greve1987vbluw,bond2025discovery,fesen2025deep,boumis2022deep,shapiro1987cosmological}. In contrast to diffuse nebulae primarily ionized by photoionization (e.g., PNe, H\,\textsc{ii} regions), SNRs are dominated by high-energy shock-excited ionization and emission, which can raise local electron temperatures above $10^5$ K. After the explosion, supersonic ejecta with high kinetic energy interact with the ambient ISM to drive shocks and produce a shock-heated plasma shell. After decelerating below $\sim$ 200 km s$^{-1}$ it becomes radiative, then rapid cooling and compression produce bright optically or UV narrow filaments characterized by strong H$\alpha$ and forbidden-line emission from N, O, and S. Previous studies \citep{makarenko2023supernova} have shown [S\,\textsc{ii}] and [N\,\textsc{ii}] emissions typically originate from the thin shell surrounding the SNR, while [O\,\textsc{iii}] and H$\alpha$ arise from the hot, volume-filling gas within the SNR bubble. Meanwhile the ratio [S\,\textsc{ii}]/H$\alpha > 0.45$ is commonly adopted to separate SNRs from H\,\textsc{ii} regions and other photoionized nebulae. [N\,\textsc{ii}]~$\lambda\lambda$6548,6583 is also enhanced in shock-excited gas. Under conditions of lower shock velocity, oxygen ionization can remain restricted to the [O\,\textsc{iii}] state. In H\,\textsc{ii} regions, however, the intensity of the [O\,\textsc{iii}] spectral line acts primarily as a tracer for the effective temperature of the ionizing stellar source. 
Large, evolved PNe can be optically similar to small SNRs but generally exhibit a central star and weaker [S\,\textsc{ii}] emission. Likewise, large ionized bubbles or H\,\textsc{ii} regions can show relatively strong [O\,\textsc{iii}] under particular conditions, though their driving sources are usually identifiable. Consequently, the presence and relative intensity of [O\,\textsc{iii}] provide valuable diagnostics for distinguishing shock versus photoionization conditions.

Complete spectroscopic coverage is essential but challenging for studying large-scale nebulae. Wide-field narrow-band imaging surveys target specific emission-line distributions of universal objects using narrow-band filters on large-field instruments, enabling the discovery of unidentified objects. Unlike broadband surveys, which provide integrated flux over wide spectral ranges, narrow-band surveys utilize filters with bandwidths of only a few nanometers to isolate key emission lines such as H$\alpha$, [O\,\textsc{iii}]~$\lambda\lambda$4959,5007, [N\,\textsc{ii}]~$\lambda\lambda$6583,6548 and [S\,\textsc{ii}]~$\lambda\lambda$6716,6732. This method significantly suppresses background airglow and light pollution, achieving detection depths of 25--30~mag/arcsec$^2$. Such surveys are crucial for identifying evolved planetary nebula shells, faint supernova remnants, large-scale ionized hydrogen bubbles, and extremely low-surface-brightness structures within and beyond the Milky Way. The first all-sky narrowband imaging efforts began in the late 20th century with the Virginia Tech Spectral-Line Survey (VTSS, \cite{dennison1999virginia}) and the Southern H-Alpha Sky Survey Atlas (SHASSA, \cite{gaustad2001robotic}). Subsequent surveys, including the INT Photometric H$\alpha$ Survey (IPHAS, \cite{drew2005int}) and the VST Photometric H$\alpha$ Survey (VPHAS+, \cite{drew2014vst}), significantly expanded the statistical samples of planetary nebulae, emission-line stars, and supernova remnants across the Galactic plane. With the widespread adoption of CCD and CMOS imaging technology, amateur astronomers have played an increasingly important role in deep narrowband imaging. Using small-aperture telescopes (5--30~cm) with commercial narrowband filters, amateur observers now achieve high signal-to-noise ratio (S/N) images through long-exposure techniques. Beyond traditional H$\alpha$ imaging, recent efforts have extended to [O\,\textsc{iii}] and [S\,\textsc{ii}] emission lines, providing multiwavelength insights into the interstellar medium and galaxy evolution. 
In recent years, the Large Sky Area Multi-Object Fiber Spectroscopic Telescope  \citep[LAMOST][]{wang1996special,su2004active,cui2012large,zhao2012lamost,luo2012data} Medium-Resolution Spectroscopic Survey of Nebulae \citep[MRS-N][]{wu2021lamost,zhang2021lamost,ren2021radial,wu2022data} has supplied astronomers with spectroscopic data for studying large-scale Galactic nebulae, alleviating the long-standing scarcity of nebular spectra.
 
Large-scale [O\,\textsc{iii}] emission structures in the Milky Way and external galaxies have been relatively rare in optical studies due to their large angular extent and low surface brightness, yet they are important diagnostics for identifying unknown emission nebulae. Deep survey has shown that in many large remnants the [O\,\textsc{iii}]~$\lambda$5007 emission traces shock structures that are more extended than those seen in H$\alpha$ \citep{fesen2020g107,fesen2024deep}, indicating that [O\,\textsc{iii}] imaging is essential to fully capture SNR shock emission, and thereby aids the detection and characterization of potential large-scale supernova remnants. The object studied in this paper, named \textit{Silk Nebula}, was first identified by amateur astronomers using narrow-band [O\,\textsc{iii}] emission line imaging. It is located near Gemini (RA = 116.68$^\circ$, Dec = +16.42$^\circ$) and spans a field of roughly 1$^\circ$ $\times$ 10$^\circ$, appearing as an elongated diffuse emission structure that suggests the presence of extended, highly ionized gas. 
The entire deep color composite image covering the \textit{Silk Nebula}, combining H$\alpha$ (red) and [O\,\textsc{iii}] (blue) emission, obtained by the SWONS (Super Wide-Field [O\,\textsc{iii}] Narrowband Survey) team, is presented in Fig~\ref{fig:OIIIMAP.png}a.
The southern sector exhibits elevated surface brightness and a pronounced, multilayered filamentary morphology, reminiscent of a supernova remnant, with measurable curvature. Corresponding structures are essentially absent in the [S\,\textsc{ii}] and H$\alpha$ bands. 
% Photometric calibration of the [O\,\textsc{iii}] image was performed using $V$ band magnitudes of nearby field stars, yielding a surface brightness of $\sim$$24.3\ \mathrm{mag\ arcsec^{-2}}$. 
Toward the north, the nebula displays a faint, arcuate band of similar curvature, which remains marginally detectable even in deep exposures.

Classification of large gaseous nebulae is initially guided by morphology (e.g., presence of a central star or a shell, and overall geometry—spherical/elliptical versus diffuse/irregular) and is subsequently refined using spectroscopic diagnostics. Multi‑wavelength emission‑line analysis yields key physical parameters (electron temperature $T_{\mathrm{e}}$, electron density $n_{\mathrm{e}}$, ionic/elemental abundances, shock velocity, and kinematics). Empirical criteria—such as [S\,\textsc{ii}]~$\lambda\lambda6716,6732/\mathrm{H}\alpha \gtrsim 0.45$ for supernova remnants versus $\lesssim 0.2$ for typical H\,\textsc{ii} regions—together with diagnostic diagrams (e.g., classical BPT diagrams and SMB‑type diagnostics \citep{kopsacheili2020diagnostic,sabin2013new}) effectively distinguish shock‑excited nebulae from photoionized regions. 
This raises the question: Is the \textit{Silk Nebula} a shock-dominated SNR or a photoionized structure? The detection of extended [O\,\textsc{iii}] emission on large angular scales is frequently associated with SNRs. However, robust observational evidence is required to substantiate this interpretation. This motivates a comprehensive spectroscopic analysis to derive the nebula's physical parameters and determine its classification.

In this paper, we describe the observational procedures, and data products for this study, with the primary datasets obtained from two major wide-field survey projects in Section~\ref{sect:obser}. 
Section~\ref{sect:resul} presents the spectroscopic data results, systematically listing key spectral line velocities, ratios, and full widths at half-maximum (FWHMs). Based on the results in Section~\ref{sect:resul}, classification diagnostics, kinematic distance, metal abundance, electron temperature and density are discussed in Section~\ref{sect:discu}. Our conclusion is presented in Section~\ref{sect:concl}, followed by a brief overview of the study's limitations and future research directions.
%---------------------------------------------------------------
   \begin{figure*}[h!]
   \centering
   \includegraphics[width=0.9\textwidth]{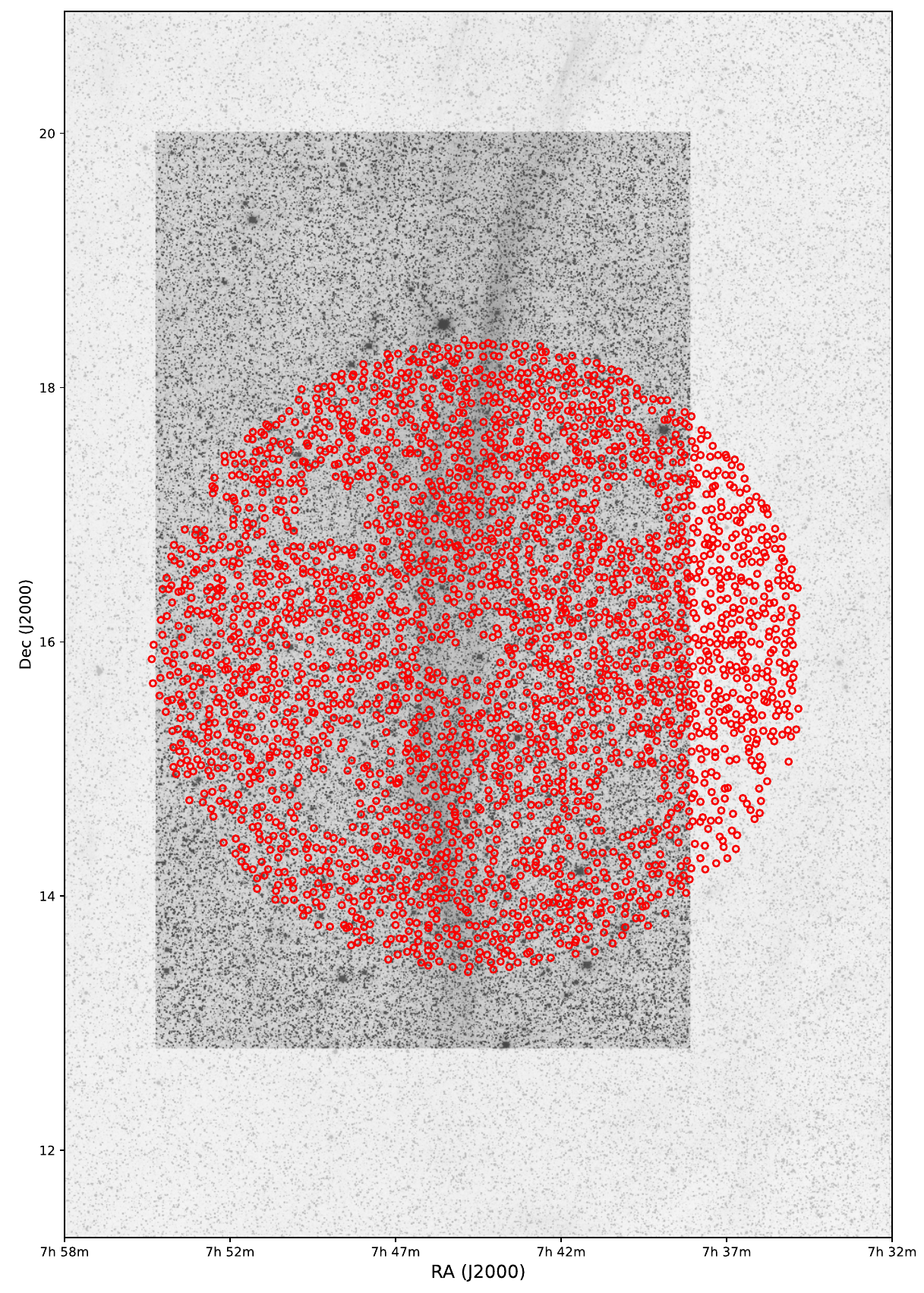}
    \caption{Fiber distribution of test sky area. Red circles are coordinates of 4000 fibers. The grayscale background shows the H$\alpha$ filter image, the colored overlay represents the [O\,\textsc{iii}] filter image. Together to fully illustrate the LAMOST field of view coverage and the location of the \textit{Silk Nebula}.}
   \label{fig:map150.png}
   \end{figure*} 
%---------------------------------------------------------------

\section{Observations}
\label{sect:obser}
\subsection{The SWONS Project}
To conduct a large-scale, low-surface-brightness [O\,\textsc{iii}] imaging survey across the Milky Way and search for ancient supernova remnants and diffuse ionized structures, a Chinese amateur astronomy team initiated the Super Wide-Field [O\,\textsc{iii}] Narrowband Survey (SWONS) in early 2023. The survey instrument consists of a ZHONGYI Optic SpeedMaster 50~mm F0.95 lens coupled with a QHY 1920M scientific-grade camera. The lens aperture was modified to incorporate an [O\,\textsc{iii}] narrowband filter (CWL = 500.7~nm, FWHM = 6.0~nm) while retaining the ultra-fast F0.95 light-gathering capability. This system covers approximately $26^\circ \times 16^\circ$ ($\sim$400~deg$^2$ per exposure) with a pixel scale of $\sim$50$''$/pixel, sufficient for detecting structures larger than $1^\circ$ in diameter. The instrument is installed at the Gemini Observatory near Lijiang Gaomeigu Station (The Sky Quality Meter (SQM) yielded a reading of $\sim$21.5~mag/arcsec$^2$).

SWONS prioritizes underobserved regions, employing a $4\times4$ mosaic imaging strategy with 20\% overlap between adjacent fields. Each tile consists of $\geq$ 20 $\times$ 180~s exposures, totaling $\geq$ 16~hours of integration per region. Initial tests revealed that artificial flat fields (dome/twilight) introduced significant vignetting, so moonless midnight sky flats were adopted for monthly calibration. Dark and bias frames were also updated monthly. After calibration, individual frames were combined into a $>$ 5000~deg$^2$ [O\,\textsc{iii}] mosaic. Foreground stars were removed, and images were nonlinearly stretched and inverted for candidate selection.

In November 2023, a candidate north-south filamentary [O\,\textsc{iii}] structure was detected in the Gemini winter field. To exclude optical artifacts (e.g., lens flares or mosaic seams), the same system reobserved the region with an additional 1-hour exposure. Follow-up observations with a 180~mm (f/2.8) reflector and a QHY 600M cooled camera (total exposure: 36~hours, resolution: $\sim$1.55$''$/pixel) confirmed the object's morphology and boundaries. The team designated this structure the \textit{Silk Nebula}.
\subsection{The LAMOST MRS-N observation of \textit{Silk Nebula}}
LAMOST, also known as the Guoshoujing Telescope, has an effective aperture of 4\,m, a large field of view of 20 square degrees and 4000 fibers, making it very suitable  for spectroscopic observation of Galactic large-scale nebulae. The medium-resolution survey \citep[MRS][]{Liu2020} has a spectral resolution of R $\sim$ 7500, with wavelength coverage of the blue arm 4950--5350\AA\, and the red arm 6300--6800\AA. 
As a subproject of MRS, MRS-N primarily observes the four emission lines (including H$\alpha$, [N\,\textsc{ii}]~$\lambda\lambda$6548,6583, [S\,\textsc{ii}]~$\lambda\lambda$6716,6732 and [O\,\textsc{iii}]~$\lambda\lambda$4959,5007) from various nebulae in the Milky Way.
The spectra analyzed here are drawn from the LAMOST MRS-N program. Since September 2018, MRS-N has mapped $\sim$1200 square degrees of the northern Galactic plane ($105^\circ \leq l \leq 215^\circ$, $-5^\circ \leq b \leq 5^\circ$), with standard field exposures of $3 \times 900$\,s, making it one of the largest nebular spectral surveys to date.
MRS-N targets include H\,\textsc{ii} regions, SNRs, DIG, some large scale Herbig--Haro objects and PNe. 

On 2024 Feb.\ 1, MRS-N observed the \textit{Silk Nebula} region. The field center is HIP\,37797 and individual fiber pointings are indicated by red circles in Fig~\ref{fig:map150.png}. The field was observed three times with exposures of 900~s per visit, yielding 2,684 usable spectra (defined as red-arm nebular emission lines with S/N $>3$). Of these, 279 spectra show relatively strong [O\,\textsc{iii}] emission and are spatially coincident with the nebular contours. The seeing during the observations was $\sim$ 5$''$; although relatively large, this had negligible impact on the nebular measurements. Since nebular spectra are more complex than typical targets, MRS-N employs a dedicated reduction pipeline. Starting from the LAMOST 1D \citep{luo2015first} products (bias/dark subtraction, flat-fielding, spectral extraction, and first-pass wavelength calibration), the MRS-N pipeline performs cosmic-ray rejection, merging of sub-exposures, fitting of sky emission lines, sky-background subtraction (optional), secondary wavelength calibration, and measurement of nebular emission-line parameters (such as radial velocity, emission-line intensity ratios, relative flux, line width, and integrated intensity). All emission lines are modeled with Gaussian profiles. After subtracting the sky background for H$\alpha$ \citep{zhang2021lamost} and the secondary wavelength calibration procedure  \citep{ren2021radial}, the radial-velocity precision reaches $< 1.5$\,km\,s$^{-1}$, substantially improving radial velocity and kinematic diagnostics. For full details of the reduction steps and algorithms, see \citet{wu2022data}.
%---------------------------------------------------------------
   \begin{figure*}
   \centering
   \includegraphics[width=1\textwidth]{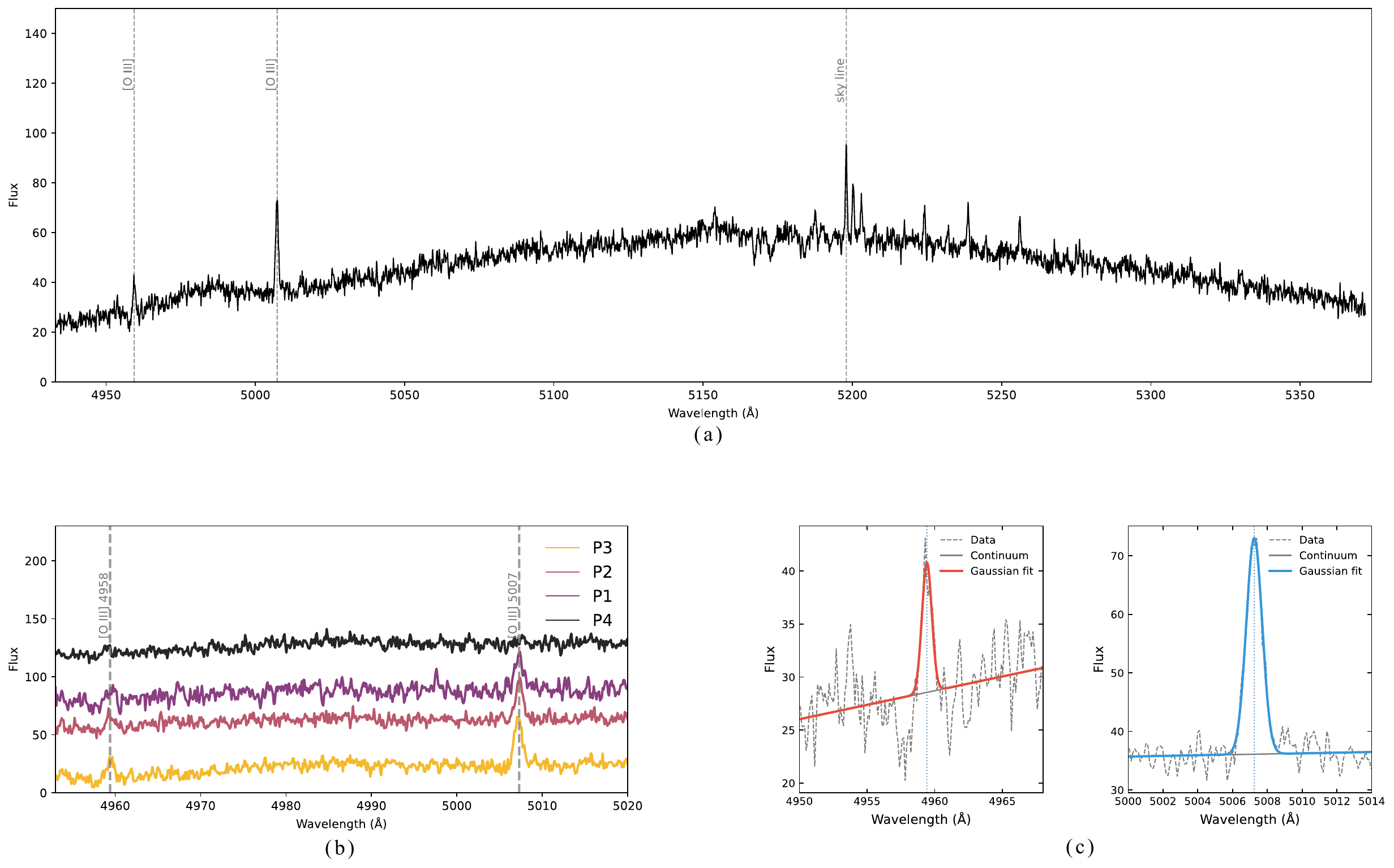}
   \caption{An example of the nebular spectra: (a) Full blue-arm spectrum. (b) Comparison of the [O\,\textsc{iii}] emission line across different regions. (c) Example of Gaussian fitting result for [O\,\textsc{iii}].}
   \label{fig:fitexample}
   \end{figure*}
%--------------------------------------------------------------------
\section{Results}
\label{sect:resul}
In this Section, we present some results obtained from imaging and spectroscopic observations.  
The \textit{Silk Nebula} was discovered by the SWONS team at the end of 2023, and its spectroscopic data were obtained on February 1, 2024. From the narrow-band images, it can be seen that the [O\,\textsc{iii}] structure has a size of about 1$^\circ$ $\times$ 10$^\circ$, and the H$\alpha$ structure is located at the northern end of the [O\,\textsc{iii}] structure, with a size of about 0.5$^\circ$ $\times$ 1$^\circ$.
The optical emission of the \textit{Silk Nebula} is detected for the first time to its full extent. 
As shown in Fig~\ref{fig:map150.png}, we present the H$\alpha$, [O\,\textsc{iii}] narrow-band image and the fiber coverage of MRS‑N. The prominent emission lines of MRS-N, [O\,\textsc{iii}]~$\lambda\lambda$4959,5007, [S\,\textsc{ii}]~$\lambda\lambda$6716,6732, [N\,\textsc{ii}]~$\lambda\lambda$6548,6583 and H$\alpha$ are measured. 
We fitted the nebular emission lines using a Gaussian function combined with a second-order polynomial. From these fits, we obtained the central wavelength, intensity, and FWHM of each emission line. Based on these parameters, we can then estimate the physical properties of the nebula, such as its radial velocity, electron temperature, electron density, and metallicity.
Since the MRS‑N spectral data lack absolute flux calibration, to eliminate the differences between different spectrographs, we selected the sky line at 5198\AA\ \citep{hanuschik2003flux} in the blue arm and the sky line at 6553 \AA\ in the red arm, respectively, to calculate the relative flux of the four nebular lines.
To facilitate comparison across different regions, the nebula was divided into three regions (P1, P2, and P3) based on the intensity morphology of [O\,\textsc{iii}], with an additional background region (P4) selected to isolate diffuse emission (Fig~\ref{fig:OIIIMAP.png}b). The equatorial coordinates (J2000) of these regions are listed in Table~\ref{Table1:radec}. The spectral parameters of different regions estimated with MRS-N data are summarized in Table~\ref{Table2:lines}, and the photometric results are summarized in Table~\ref{Table4:Photometry}. 
Fig~\ref{fig:fitexample} shows the MRS-N blue-arm spectra of the \textit{Silk Nebula} (Fig~\ref{fig:fitexample}a) and the background (Fig~\ref{fig:fitexample}b). Fig~\ref{fig:fitexample}c shows the fit to the [O III] doublet. 
%---------------------------------------------------------------
\begin{deluxetable}{ccccc}
\tablecaption{Central position of the four regions\label{Table1:radec}}
\renewcommand{\arraystretch}{1.2}  % 增大行距
\tablewidth{0pt}
\tablehead{
\colhead{Region} & \colhead{P1} & \colhead{P2} & \colhead{P3} & \colhead{P4}
}
\startdata
RA [deg] & 116.65 & 116.21 & 116.45 & 115.20\\
DEC [deg] & 16.60 & 17.74 & 14.72 & 16.18
\enddata
\end{deluxetable}
%---------------------------------------------------------------
%---------------------------------------------------------------
%--------------------------------------------------------------------
%--------------------------------------------------------------------
\begin{table*}
\centering
\caption{Relative line intensities and observational parameters observed in \textit{Silk Nebula}.}
\label{Table2:lines}
\footnotesize
\setlength{\tabcolsep}{4pt}
\renewcommand{\arraystretch}{1.5} 
\begin{tabular*}{\textwidth}{@{\extracolsep{\fill}}cccccccccccccc@{}}
\hline
\multicolumn{1}{c}{   } &\multicolumn{3}{c}{ P1 } &\multicolumn{3}{c}{ P2 } &\multicolumn{3}{c}{ P3 } &\multicolumn{3}{c}{ P4 } \\

\multicolumn{1}{c}{Line\,$\mathring{A}$} 
&\multicolumn{1}{c}{S/N\textsuperscript{a}} &\multicolumn{1}{c}{I\textsuperscript{b}} &\multicolumn{1}{c}{FWHM\textsuperscript{c}} 
&\multicolumn{1}{c}{S/N} &\multicolumn{1}{c}{I} &\multicolumn{1}{c}{FWHM}
&\multicolumn{1}{c}{S/N} &\multicolumn{1}{c}{I} &\multicolumn{1}{c}{FWHM}
&\multicolumn{1}{c}{S/N} &\multicolumn{1}{c}{I} &\multicolumn{1}{c}{FWHM}\\ 
\hline

% 第1行：[S II]6717，FWHM 用 multirow 向下合并2行
\multicolumn{1}{c}{[S\,\textsc{ii}]6716} 
&\multicolumn{1}{c}{55} &\multicolumn{1}{c}{0.14$\pm$0.02} &\multicolumn{1}{c}{\multirow{2}{*}{29.0$\pm$6.1}} 
&\multicolumn{1}{c}{52} &\multicolumn{1}{c}{0.17$\pm$0.02} &\multicolumn{1}{c}{\multirow{2}{*}{31.4$\pm$6.4}}
&\multicolumn{1}{c}{65} &\multicolumn{1}{c}{0.11$\pm$0.02} &\multicolumn{1}{c}{\multirow{2}{*}{30.1$\pm$6.1}}
&\multicolumn{1}{c}{44} &\multicolumn{1}{c}{0.10$\pm$0.02} &\multicolumn{1}{c}{\multirow{2}{*}{32.9$\pm$6.4}}\\

% 第2行：[S II]6731，FWHM 单元格留空
\multicolumn{1}{c}{[S\,\textsc{ii}]6732} 
&\multicolumn{1}{c}{55} &\multicolumn{1}{c}{0.13$\pm$0.02} &\multicolumn{1}{c}{} 
&\multicolumn{1}{c}{52} &\multicolumn{1}{c}{0.15$\pm$0.02} &\multicolumn{1}{c}{}
&\multicolumn{1}{c}{65} &\multicolumn{1}{c}{0.10$\pm$0.02} &\multicolumn{1}{c}{}
&\multicolumn{1}{c}{44} &\multicolumn{1}{c}{0.10$\pm$0.02} &\multicolumn{1}{c}{}\\

% 第3行：[N II]6548，FWHM 用 multirow 向下合并2行
\multicolumn{1}{c}{[N\,\textsc{ii}]6548} 
&\multicolumn{1}{c}{47} &\multicolumn{1}{c}{0.09$\pm$0.02} &\multicolumn{1}{c}{\multirow{2}{*}{27.0$\pm$4.6}} 
&\multicolumn{1}{c}{48} &\multicolumn{1}{c}{0.06$\pm$0.02} &\multicolumn{1}{c}{\multirow{2}{*}{26.1$\pm$4.7}}
&\multicolumn{1}{c}{53} &\multicolumn{1}{c}{0.07$\pm$0.02} &\multicolumn{1}{c}{\multirow{2}{*}{30.0$\pm$6.0}}
&\multicolumn{1}{c}{54} &\multicolumn{1}{c}{0.08$\pm$0.02} &\multicolumn{1}{c}{\multirow{2}{*}{26.9$\pm$4.8}}\\

% 第4行：[N II]6584，FWHM 单元格留空
\multicolumn{1}{c}{[N\,\textsc{ii}]6583} 
&\multicolumn{1}{c}{47} &\multicolumn{1}{c}{0.24$\pm$0.02} &\multicolumn{1}{c}{} 
&\multicolumn{1}{c}{48} &\multicolumn{1}{c}{0.26$\pm$0.03} &\multicolumn{1}{c}{}
&\multicolumn{1}{c}{53} &\multicolumn{1}{c}{0.20$\pm$0.02} &\multicolumn{1}{c}{}
&\multicolumn{1}{c}{54} &\multicolumn{1}{c}{0.20$\pm$0.02} &\multicolumn{1}{c}{}\\

% 第5行：Hα，无合并
\multicolumn{1}{c}{H$\alpha$} 
&\multicolumn{1}{c}{50} &\multicolumn{1}{c}{0.30$\pm$0.02} &\multicolumn{1}{c}{31.9$\pm$4.0} 
&\multicolumn{1}{c}{45} &\multicolumn{1}{c}{0.37$\pm$0.02} &\multicolumn{1}{c}{31.0$\pm$3.0}
&\multicolumn{1}{c}{53} &\multicolumn{1}{c}{0.26$\pm$0.02} &\multicolumn{1}{c}{32.2$\pm$4.2}
&\multicolumn{1}{c}{45} &\multicolumn{1}{c}{0.25$\pm$0.02} &\multicolumn{1}{c}{32.4$\pm$3.4}\\

% 第6行：[O III]5007，FWHM 用 multirow 向下合并2行
\multicolumn{1}{c}{[O\,\textsc{iii}]5007} 
&\multicolumn{1}{c}{6.1} &\multicolumn{1}{c}{1.67} &\multicolumn{1}{c}{\multirow{2}{*}{46.9}} 
&\multicolumn{1}{c}{6.1} &\multicolumn{1}{c}{1.83} &\multicolumn{1}{c}{\multirow{2}{*}{48.9}}
&\multicolumn{1}{c}{6.6} &\multicolumn{1}{c}{1.90} &\multicolumn{1}{c}{\multirow{2}{*}{43.1}}
&\multicolumn{1}{c}{-}   &\multicolumn{1}{c}{-}    &\multicolumn{1}{c}{\multirow{2}{*}{-}}\\

% 第7行：[O III]4958，FWHM 单元格留空
\multicolumn{1}{c}{[O\,\textsc{iii}]4959} 
&\multicolumn{1}{c}{1.8} &\multicolumn{1}{c}{0.73} &\multicolumn{1}{c}{} 
&\multicolumn{1}{c}{2.0} &\multicolumn{1}{c}{0.75} &\multicolumn{1}{c}{}
&\multicolumn{1}{c}{2.3} &\multicolumn{1}{c}{0.75} &\multicolumn{1}{c}{}
&\multicolumn{1}{c}{-}   &\multicolumn{1}{c}{-}    &\multicolumn{1}{c}{}\\

% 第8行
\multicolumn{1}{c}{[S\,\textsc{ii}]/H$\alpha$} &\multicolumn{3}{c}{0.93} &\multicolumn{3}{c}{0.88} &\multicolumn{3}{c}{0.83} &\multicolumn{3}{c}{0.78} \\

% 第9行
\multicolumn{1}{c}{[N\,\textsc{ii}]/H$\alpha$} &\multicolumn{3}{c}{1.11} &\multicolumn{3}{c}{0.88} &\multicolumn{3}{c}{1.02} &\multicolumn{3}{c}{0.95} \\

% 第10行
\multicolumn{1}{c}{[S\,\textsc{ii}]6716/6732} &\multicolumn{3}{c}{1.10} &\multicolumn{3}{c}{1.14} &\multicolumn{3}{c}{1.19} &\multicolumn{3}{c}{1.03} \\
\hline

\end{tabular*}
\vspace{0.5em}

\begin{minipage}{\textwidth}  
\scriptsize
$^a$ The signal to noise ratio of the observed emission lines\\  
$^b$ The emission lines relative intensity, expressed as the ratio to the sky line intensity\\
$^c$ FWHM of emission lines, in units of km\,s$^{-1}$, have been corrected for instrumental broadening\\
\end{minipage}
\end{table*}

   \subsection{Photometry}
   Owing to the nebula's diffuse nature and low surface brightness, we manually selected star‑free subregions within P1–P3 and adjacent nebula‑free background regions in the narrow‑band images to ensure reliable photometry. Source detection and initial photometry were performed with SExtractor, and aperture photometry was carried out for all isolated, unsaturated, non‑variable field stars in the images. The derived aperture magnitudes were cross‑matched with the Gaia and SDSS catalogues to determine the photometric zeropoint for each region, thereby completing the photometric calibration. To minimise systematic uncertainties, the Gaia (G, BP, RP) and SDSS (g, r) magnitudes were first transformed into a common photometric system following the method of \citet{jordi2010gaia}. The SExtractor counts were then converted to surface brightness using the derived zeropoints and exposure times. All uncertainties—including statistical measurement errors, background‑estimation uncertainties, and zeropoint calibration errors—were propagated through standard error propagation and are listed alongside the results in Table~\ref{Table4:Photometry}. Each nebular region is paired with its corresponding background region, and both share the same photometric zeropoint. 
   % \textcolor{purple}{
   Although P1 appears the faintest in the narrowband image, it yields the highest measured surface brightness. This discrepancy likely arises because the dense stellar field in this region contaminates the nebular photometry with stellar light.
   % }
   As shown in Table~\ref{Table4:Photometry}, the low surface brightness of the nebula indicates that the emitting gas is tenuous and of low density.
\begin{table}
\centering 
\caption{Photometric results of \textit{Silk Nebula}}
\label{Table4:Photometry}
\renewcommand{\arraystretch}{1.5} 
\begin{tabular}{c c c c c}
\hline
Field  & Silk Nebula\textsuperscript{*}  & Zeropoint\textsuperscript{*} \\
\hline
P1 &  $23.60 \pm 0.028$ & $11.23$ \\
\hline
P2 &  $23.87 \pm 0.035$ & $11.32$ \\
\hline
 P3 &  $23.85 \pm 0.039$ & $11.29$ \\
\hline
\end{tabular}
\vspace{0.5em}
\begin{minipage}{\columnwidth}  
\scriptsize
$ $\\
$ $\\
$^*$ The magnitudes listed in the table are [O\,\textsc{iii}] band magnitudes, derived by interpolation from Gaia and SDSS photometry, and are given in units of mag\,arcsec$^{-2}$\\ 

\end{minipage}
\end{table}
%--------------------------------------------------------------------
\subsection{Spectroscopy}
\subsubsection{Relative Intensity}
\begin{figure*}[h!]
   \centering
   \includegraphics[width=0.9\textwidth]{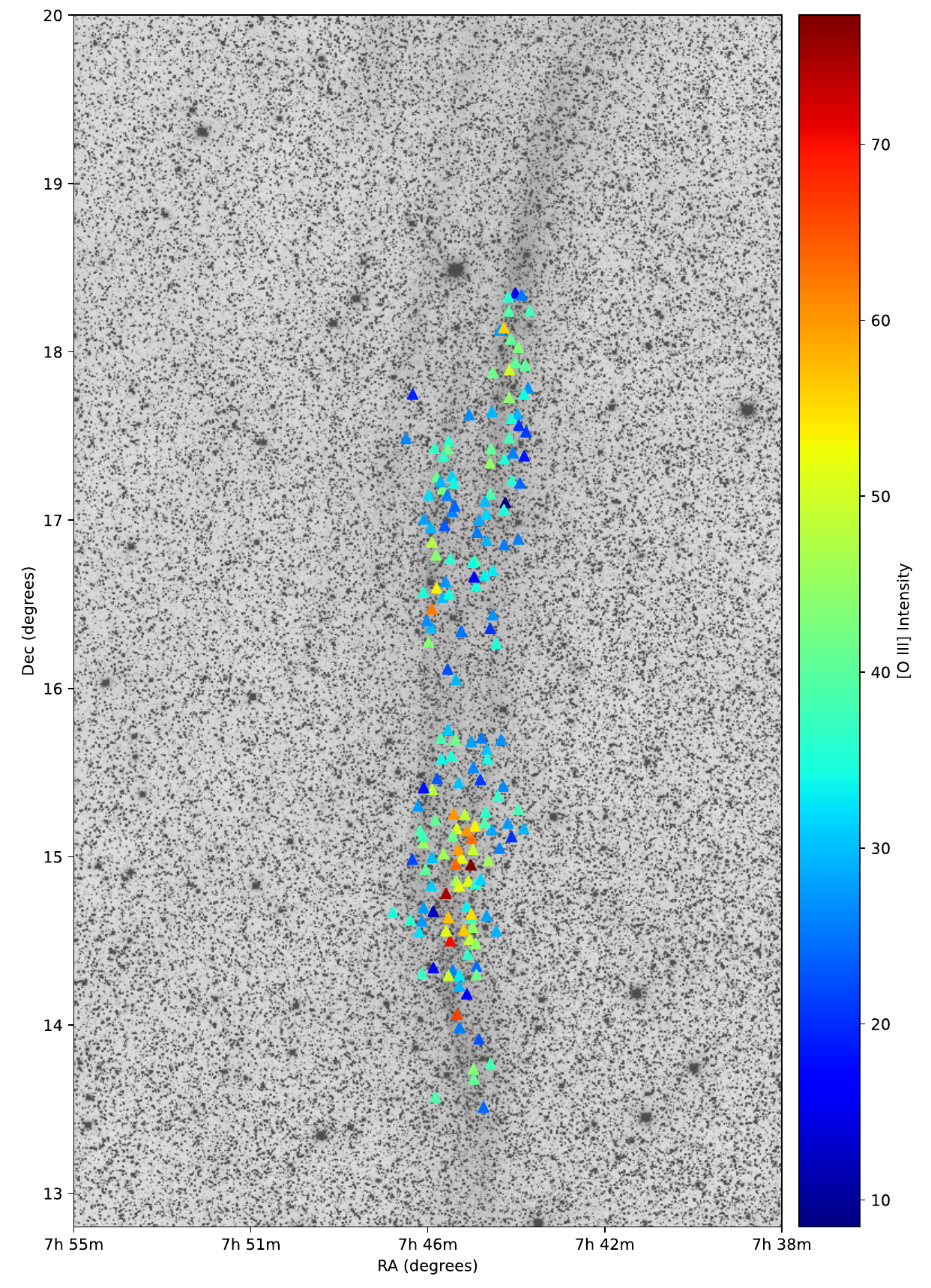}
    \caption{Relative [O\,\textsc{iii}] intensity distribution map. Plotted points mark the positions where [O\,\textsc{iii}] emission is detected, regions without significant [O\,\textsc{iii}] emission remain blank to avoid the impact of large flux‐measurement uncertainties.}
   \label{fig:OIII-RI map}
   \end{figure*}

The line intensities reported in this work are relative values, placed on a common scale using the simultaneously observed sky emission lines at 5198\AA\ (blue arm) and 6553\AA\ (red arm). Fig~\ref{fig:OIII-RI map} shows the spatial distribution of the [O\,\textsc{iii}] emission in the brighter regions of the \textit{Silk Nebula}, with intensities measured from the spectra of regions P1, P2 and P3 indicated in Fig~\ref{fig:OIIIMAP.png}. The [O\,\textsc{iii}] intensity exhibits a radial decline from the center outward and closely mirrors the morphological outline of the nebula as seen in the [O\,\textsc{iii}] narrow band image. We adopt a signal-to-noise ratio (S/N) greater than 3 as the minimum criterion for data usability. Under this threshold, the [O\,\textsc{iii}]~$\lambda$5007 emission line—which is relatively weak—exhibits an S/N of only 6.1 (Table~\ref{Table2:lines}), just above the limit required for reliable detection. In addition, region P4 was selected as a nebula-free reference, as neither the narrow band image nor the blue arm spectrum shows any discernible [O\,\textsc{iii}] emission or structure in this region (see Fig~\ref{fig:fitexample}b). The fitted parameters for the nebular lines in all four regions (P1–P4) are presented in Table~\ref{Table2:lines}. It should be noted that none of the nebular line fluxes presented here have been absolutely flux-calibrated. Consequently, the [O\,\textsc{iii}] intensity measured in the blue arm cannot be directly compared with, nor can line ratios be reliably estimated relative to, the three lines (H$\alpha$, [S\,\textsc{ii}] and [N\,\textsc{ii}]) observed in the red arm. However, the three red arm lines were obtained simultaneously with the same spectrograph and are closely spaced in wavelength. We therefore consider the effects of interstellar extinction to be negligible across this narrow range. Accordingly, the line ratios derived from these three red arm lines can be confidently used for direct physical analysis in this work. For the three red arm lines, the H$\alpha$ emission line exhibits the highest intensity, while the [N\,\textsc{ii}] line is the weakest. In addition, the [N\,\textsc{ii}]/H$\alpha$ ratio in region P2 is noticeably lower than in the other regions—barely above 0.8. 
%--------------------------------------------------------------------
%---------------------------------------------------------------
\subsubsection{The FWHM}
   The width of an emission line in a nebular spectrum encodes fundamental information about the kinematics and physical state of the ionized gas along the line of sight. Intrinsic line broadening arises from a combination of thermal Doppler motions—which scale with the square root of the gas temperature—and non-thermal processes such as small-scale turbulence, bulk flows, and the expansion of nebular shells. Because thermal broadening depends on the atomic mass of the emitting ion, comparing the line widths of light elements (e.g., hydrogen) with those of heavier species (e.g., oxygen or sulfur) allows one to disentangle the thermal contribution from the non-thermal turbulent component. Furthermore, deviations from a purely Gaussian profile, or the presence of line splitting, can reveal unresolved velocity structure, including expanding bubbles, the superposition of multiple structures, or shock fronts. To extract these physical insights reliably, it is essential to distinguish the intrinsic line width from the observed line width by accurately accounting for instrumental broadening, which otherwise artificially inflates the measured FWHM. 
   
   To recover the intrinsic line width, the instrumental broadening must be deconvolved or subtracted from the observed FWHM. \cite{zhang2021lamost} compared the line dispersions of arc-lamp lines with that of the 6553\AA\ skyline in LAMOST MRS-N spectra and found them to be fully consistent. Both arc lines and sky lines are commonly used to characterize instrumental broadening in spectroscopic observations. Since the instrumental profile may vary from fiber to fiber, we corrected the instrumental broadening for each spectrum using its corresponding sky line. Following the method of \cite{reynolds1977faint}, we applied the equation:
   \begin{equation}
          \mathrm{FWHM}^2_{\mathrm{obs}} = \mathrm{FWHM}^2_{\mathrm{real}} + \mathrm{FWHM}^2_{\mathrm{inst}}\label{equ:FWHMobs}
   \end{equation}
   where $\mathrm{FWHM}_{\mathrm{obs}}$ is the observed line width, $\mathrm{FWHM}_{\mathrm{inst}}$ denotes the instrumental broadening, and $\mathrm{FWHM}_{\mathrm{real}}$ represents the intrinsic FWHM of the emission line. To ensure data reliability, we applied the criterion \citep{li2022lamost} $\mathrm{FWHM}_{\mathrm{obs}} - \mathrm{errFWHM}_{\mathrm{obs}} > \mathrm{FWHM}_{\mathrm{inst}} + \mathrm{errFWHM}_{\mathrm{inst}}$ (here $\mathrm{errFWHM}_{\mathrm{obs}}$ and $\mathrm{errFWHM}_{\mathrm{inst}}$ denote the uncertainties in the observed and instrumental FWHM, respectively). 
   All FWHM values reported in Table~\ref{Table2:lines} have been corrected for instrumental broadening and thus represent the intrinsic line widths. From these FWHM measurements, key physical properties of the nebula—such as temperature and shock velocity—can be inferred. More detailed can be seen in Section~\ref{subsec:TD}. The data in Table~\ref{Table2:lines} show the line widths in regions P1, P2, P3, and P4 are remarkably similar. Taken together with the discussion in Section~\ref{sect:discu}, this suggests additional insights into the nebular structure.
%---------------------------------------------------------------
\subsubsection{Radial Velocity}
%---------------------------------------------------------------
   \begin{figure*}
   \centering
   \includegraphics[width=1.0\textwidth]{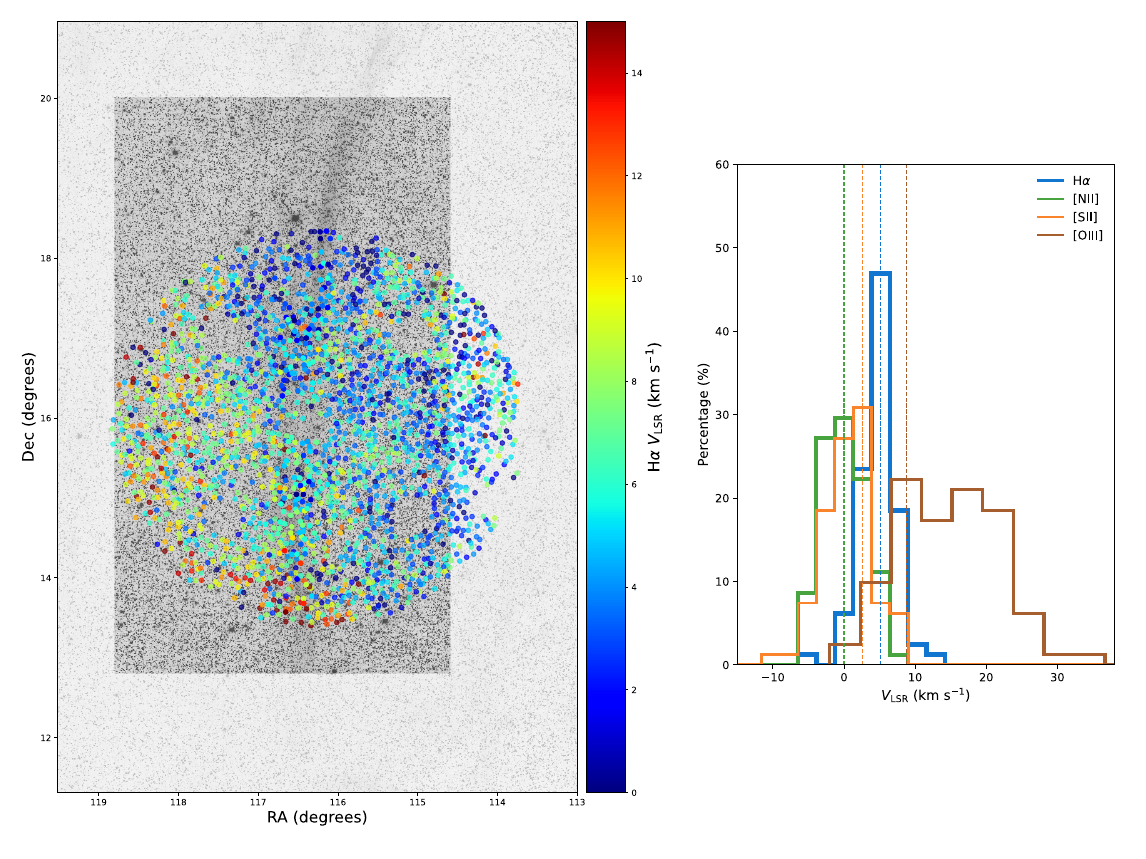}
   \caption{Two-dimensional spatial distributions (left panel) of $V_{\rm LSR}$ (in km\,s$^{-1}$) measured with H$\alpha$ and the histograms (right panel) of $V_{\rm LSR}$ of the H$\alpha$, [N\,\textsc{ii}], [S\,\textsc{ii}] and [O\,\textsc{iii}] emission lines.}
   \label{fig:RV}
   \end{figure*}
%---------------------------------------------------------------
The MRS-N spectra have been corrected for heliocentric motion using the LAMOST official pipeline. Consequently, the radial velocities measured from the emission lines represent heliocentric velocities ($V_{\rm helio}$, in km\,s$^{-1}$). The $V_{\rm helio}$ of P1, P2 and P3 traced by [O\,\textsc{iii}] with uncertainties of 3-4 km\,s$^{-1}$, while P4 is better characterized by the stronger H$\alpha$ line, yielding a smaller uncertainty of $\sim$ 1.4 km\,s$^{-1}$. $V_{\rm helio}$ analysis reveals a discrepancy of $\sim$ 5$-$15 km\,s$^{-1}$ between the [O\,\textsc{iii}] lines and the [N\,\textsc{ii}], [S\,\textsc{ii}], and H$\alpha$.  The local standard of rest (LSR)
velocity ($V_{\rm LSR}$) was obtained by correcting the heliocentric velocities computed above for the solar peculiar motion. We adopt a solar velocity of $\simeq$ 20 km\,s$^{-1}$ directed toward $\alpha = 18^{\mathrm{h}}03^{\mathrm{m}}50.29^{\mathrm{s}}$, $\delta$ = +30$^{\circ}$ 00$^{\prime}$ 16.8$^{\prime\prime}$ (J2000). The histograms of $V_{\rm LSR}$ for different emission lines along the line of sight to the nebula and two-dimensional spatial distributions of $V_{\rm LSR}$ measured with H$\alpha$ are revealed in Fig~\ref{fig:RV}. The data and associated uncertainties of RVs and $V_{\rm LSR}$ for the different regions are presented in Table~\ref{Table3:V}.  

\subsection{Distance} \label{subsec:distance}
A major problem for determining the physical properties of SNRs or nebulae is the uncertainty in their distances, which are needed to transform angular sizes into physical sizes and proper motions into velocities, but it is also important for determining the three-dimensional spatial distribution in the Galaxy\citep{vink2020physics}. While numerous distance measurement methods are available, the options are substantially reduced for large-scale nebulae. 
% Deriving spectrophotometric distance needs to locate the exciting star(s) in the optical or near-infrared and assign the spectral type. 
Unlike H\,\textsc{ii} regions, where spectrophotometric distances can be derived by identifying the massive exciting star(s) and determining their spectral types, SNRs are powered by shock-heated interstellar gas and lack such ionizing sources. Therefore, spectrophotometric distance estimation is not applicable to the \textit{Silk Nebula} if it is indeed a SNR. 
Spectrophotometric distances are also susceptible to extinction, which is poorly constrained along this low-reddening sightline. Maser parallax distances, meanwhile, require VLBI measurements of bright maser sources associated with massive star-forming regions \citep{anderson2012green} and are therefore inapplicable to this object. Kinematic distances offer a more suitable alternative, though they are subject to systematic uncertainties from the rotation curve, non-circular motions, and velocity crowding.
Integration with extinction–distance method\citep{wang2020distances,chen2025distances} offers a more reliable determination of the nebular distance.

\begin{figure*}
   \centering
   \includegraphics[width=1.0\textwidth]{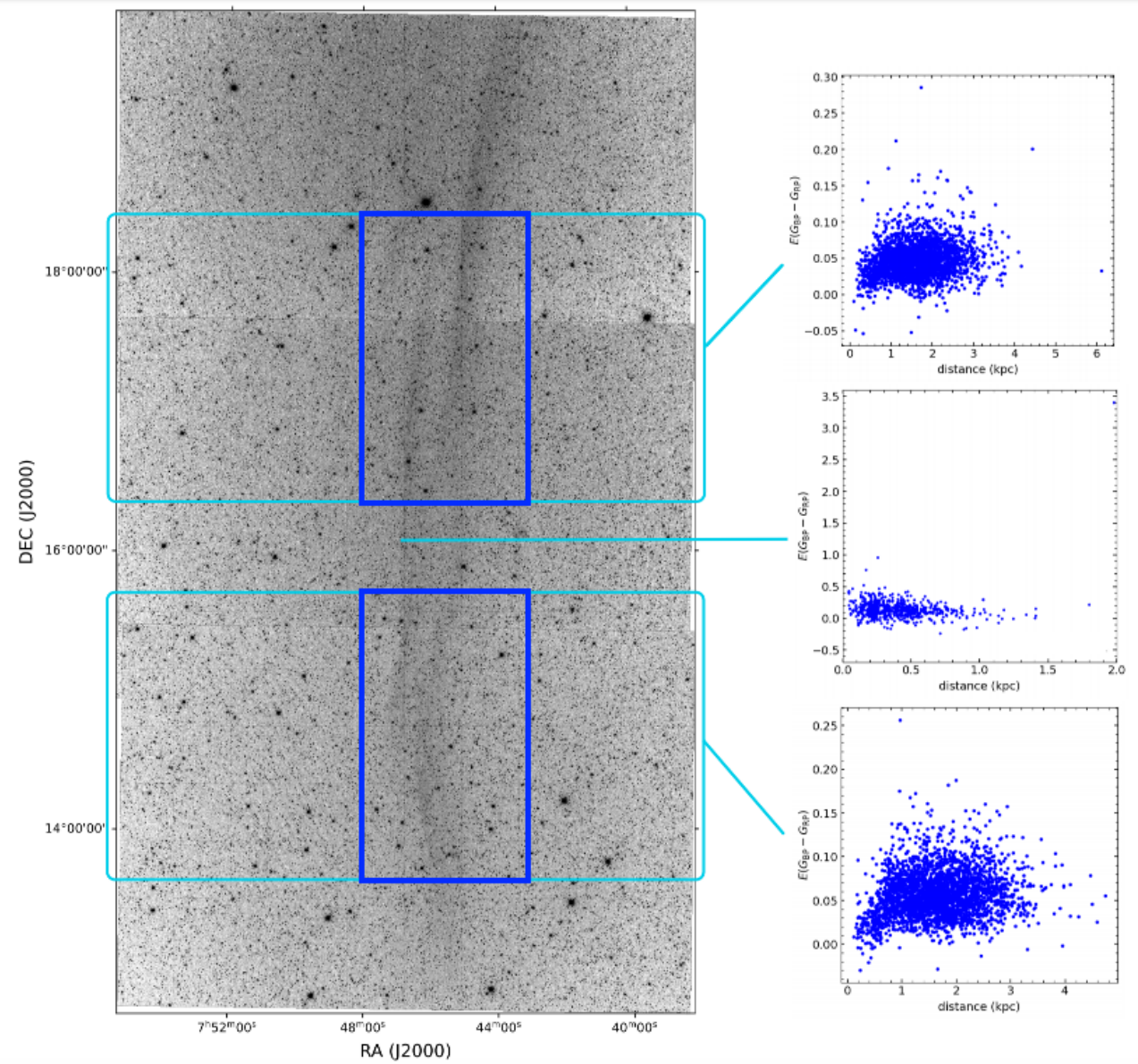}
    \caption{Intrinsic color versus distance for stars projected on the nebula  and for all stars along the line of sight . Both datasets show consistent behavior and no significant extinction jump with distance.}
   \label{fig:starE}
   \end{figure*}

%---------------------------------------------------------------
%\begin{figure*}
    %\centering
    %\includegraphics[width=0.8\textwidth]{curve and map.pdf}
    %\caption{(a) Extinction maps shown at distance slices of 300, 1000 and 5000 pc, centered on the nebula (orange circle).
   % (b) Extinction as a function of distance modulus for the line of sight through the nebula (best-fit profile shown in bold, individual samples in thin lines). The extinction profiles measured in different subregions are similar and indicate very weak reddening at all distances; the nearest distance range is poorly constrained because of the paucity of foreground stars. }
    %\label{fig:curve map}
%\end{figure*}
%---------------------------------------------------------------
We analyzed the extinction data and derived kinematic distance. By using the 3D dust map based on Gaia, Pan-STARRS1, and 2MASS \citep{green20193d}, we detect no significant extinction features across the distance slices at 300, 1000, and 5000 pc. The similar trends of extinction versus distance modulus across various subregional lines of sight suggest the absence of a significant extinction jump with distance, although the nearby range is poorly constrained due to the scarcity of stars. Consequently, we infer that the extinction along this line of sight is remarkably low, rendering the dust extinction data insufficient to provide meaningful constraints on the distance to the nebula. 
% Therefore, the line of sight exhibits very low extinction, and the dust-extinction information cannot be used to place a meaningful constraint on the nebular distance. 
The same conclusion is supported by the intrinsic color indices of stars projected on the nebula and along the line of sight (Fig~\ref{fig:starE}).

The kinematic distances are calculated using the Python package downloaded from \url{https://github.com/tvwenger/kd} \citep{wenger2018kinematic}. During the calculation, the rotation curve has been upgraded by \cite{reid2014trigonometric}. When implementing the aforementioned code, due to velocity crowding, kinematic distances are unreliable in the direction of the Galactic Center (GC; $\ell$ = 0$^{\circ}$) and the Galactic Anti-center (GAC; $\ell$ = 180$^{\circ}$). 
Owing to the negligible contribution of circular rotation to $V_{\rm LSR}$ along these sightlines, 
 % Because the line-of-sight velocity from circular rotation approaches zero in these directions, 
 objects lying within 15$^{\circ}$ of the GC and 20$^{\circ}$ of the GAC are excluded from the kinematic distance analysis (\citealp{wenger2018kinematic};\citealp{balser2015azimuthal}). 
% As the $V_{\rm LSR}$ due to circular motion tend toward zero in these directions, all objects within 15$^{\circ}$ of the GC and 20$^{\circ}$ of the GAC are expected to be excluded
We have computed kinematic distances by two methods based on the code—Traditional kinematic distance and Monte Carlo kinematic distance, and the results are summarized in Table~\ref{Table3:V}. Although located outside the excluded GAC region, the \textit{Silk Nebula} lies near its boundary, which introduces significant uncertainties into the derived kinematic distance. 
Derived $V_{\rm LSR}$ ($\sim 11$--$16\,\mathrm{km\,s^{-1}}$) and kinematic distances ($\sim 100$--$190\,\mathrm{pc}$) from [O\,\textsc{iii}] are broadly consistent across different regions, with P3 yielding a somewhat lower distance estimate (Table~\ref{Table3:V}), though the values are consistent within the Monte Carlo uncertainties. 
% Although offsets are observed between the [O\,\textsc{iii}] line and the [S\,\textsc{ii}], [N\,\textsc{ii}], and H$\alpha$ lines, the distances estimated from all four emission lines remain broadly consistent once the measurement uncertainties are taken into account.
% \textcolor{red}{However, systematic discrepancies are observed when compared with values obtained from [S\,\textsc{ii}], [N\,\textsc{ii}], and H$\alpha$ lines, which exhibit mutual consistency across regions. Considering the uncertainties inherent to nearby distances, the distances derived from all four emission lines remain mutually consistent.} 
Because the [O\,\textsc{iii}] emission is the only line that exhibits clear morphological correspondence with the nebular filamentary structure in narrowband imaging, we base our distance estimate exclusively on the [O\,\textsc{iii}] line. 
% The $\sim 15\,\mathrm{km\,s^{-1}}$ velocity offset between [O\,\textsc{iii}] and the other three emission lines raises questions about their association with a single physical structure (see Section~\ref{sect:discu} for a detailed discussion). 
% Nevertheless, t
% The close agreement between [O\,\textsc{iii}] line intensities and narrowband imaging confirms that the observed [O\,\textsc{iii}] emission indeed originates from the nebular structure. 
% Considering its angular extent exceeds 10$^{\circ}$ and adopting the kinematic distance of $\sim$ 160 pc, we derive a linear size of $\sim$ 28 pc for the nebula. Its spatial continuity and scale are rare among known [O\,\textsc{iii}] structures in the literature. 
Previous reports of large‐scale [O\,\textsc{iii}] structures have generally been limited to SNR rim‐ionization regions with angular extents of around 1$^{\circ}$ \citep{fesen2020g107,boumis2022deep}. If confirmed as a shock-dominated ionized shell of a SNR, this structure would likely represent one of the largest angular-scale [O\,\textsc{iii}] features ever recorded in previous studies. 

The $V_{\rm LSR}$ values measured from the three red arm emission lines (H$\alpha$, [N\,\textsc{ii}], and [S\,\textsc{ii}]) show a systematic offset of typically $\sim$ 10–16 km $\mathrm{s^{-1}}$ relative to the $V_{\rm LSR}$ of [O\,\textsc{iii}]. When kinematic distances are estimated independently from the three red arm lines, the resulting values differ from the [O\,\textsc{iii}]-based distance by approximately 100 pc. Given that the low-ionization emission is likely subject to contamination from other components along the line of sight -- most plausibly DIG, as discussed in detail in Section~\ref{sect:discu} -- and that no clear morphological counterpart to the red arm lines kinematic structure is apparent in our narrow-band images, 
% we interpret the red arm line velocity field as being affected by non-SNR gas. 
we adopt the distance derived from the [O\,\textsc{iii}] emission line as the most reliable estimate for the Silk Nebula throughout this work.

We therefore adopt a fiducial distance of $\sim$160~pc (the approximate median of the six [O\,\textsc{iii}]-based kinematic distance estimates; see Table~\ref{Table3:V}). At this distance, the nebula extends $\sim$28~pc in the north--south direction, with an estimated physical radius of $\sim$30~pc for the complete remnant shell.

%---------------------------------------------------------------

%--------------------------------------------------------------------
%--------------------------------------------------------------------
\begin{table*}
\centering
\caption{Velocities derived from observed emission lines in different regions}
\label{Table3:V}
\renewcommand{\arraystretch}{1.5}
\begin{tabular*}{\textwidth}{@{\extracolsep{\fill}}ccccccc@{}}
\hline
\multicolumn{1}{c}{ } &\multicolumn{1}{c}{ } &\multicolumn{1}{c}{RVs} &\multicolumn{1}{c}{$V_{\rm LSR}$} &\multicolumn{1}{c}{Distance\textsuperscript{1}} &\multicolumn{1}{c}{Distance\textsuperscript{2}} &\multicolumn{1}{c}{$V_t$} \\
\hline

\multirow{4}{*}{P1} &\multicolumn{1}{c}{[O\,\textsc{iii}]} &\multicolumn{1}{c}{26.8$\pm$4.3} &\multicolumn{1}{c}{15.5} &\multicolumn{1}{c}{$1.63^{+1.01}_{-0.84}$} &\multicolumn{1}{c}{1.79} &\multirow{4}{*}{33.1} \\

&\multicolumn{1}{c}{[S\,\textsc{ii}]} &\multicolumn{1}{c}{11.8$\pm$2.5} &\multicolumn{1}{c}{-1.0} &\multicolumn{1}{c}{$0.06^{+1.17}_{-0.05}$} &\multicolumn{1}{c}{0.16} \\

&\multicolumn{1}{c}{[N\,\textsc{ii}]} &\multicolumn{1}{c}{11.8$\pm$1.6} &\multicolumn{1}{c}{-0.9} &\multicolumn{1}{c}{$0.11^{+1.12}_{-0.10}$} &\multicolumn{1}{c}{0.18} \\

&\multicolumn{1}{c}{[H$\alpha$]} &\multicolumn{1}{c}{16.3$\pm$1.2} &\multicolumn{1}{c}{4.7} &\multicolumn{1}{c}{$0.47^{+1.18}_{-0.45}$} &\multicolumn{1}{c}{0.61}  \\

\hline

\multirow{4}{*}{P2} &\multicolumn{1}{c}{[O\,\textsc{iii}]} &\multicolumn{1}{c}{26.8$\pm$4.1} &\multicolumn{1}{c}{14.5} &\multicolumn{1}{c}{$1.58^{+1.23}_{-0.83}$} &\multicolumn{1}{c}{1.85} &\multirow{4}{*}{32.9} \\

&\multicolumn{1}{c}{[S\,\textsc{ii}]} &\multicolumn{1}{c}{12.0$\pm$1.6} &\multicolumn{1}{c}{0.5} &\multicolumn{1}{c}{$0.21^{+1.06}_{-0.20}$} &\multicolumn{1}{c}{0.17} \\

&\multicolumn{1}{c}{[N\,\textsc{ii}]} &\multicolumn{1}{c}{12.0$\pm$1.6} &\multicolumn{1}{c}{-0.5} &\multicolumn{1}{c}{$0.08^{+1.35}_{-0.07}$} &\multicolumn{1}{c}{0.17}  \\

&\multicolumn{1}{c}{[H$\alpha$]} &\multicolumn{1}{c}{17$\pm$1.2} &\multicolumn{1}{c}{4.0} &\multicolumn{1}{c}{$0.31^{+1.44}_{-0.31}$} &\multicolumn{1}{c}{0.54}  \\

\hline

\multirow{4}{*}{P3} &\multicolumn{1}{c}{[O\,\textsc{iii}]} &\multicolumn{1}{c}{23.8$\pm$3.3} &\multicolumn{1}{c}{11.2} &\multicolumn{1}{c}{$0.98^{+1.08}_{-0.69}$} &\multicolumn{1}{c}{1.20} &\multirow{4}{*}{30.6} \\

&\multicolumn{1}{c}{[S\,\textsc{ii}]} &\multicolumn{1}{c}{11.7$\pm$1.9} &\multicolumn{1}{c}{1.7} &\multicolumn{1}{c}{$0.21^{+1.12}_{-0.21}$} &\multicolumn{1}{c}{0.23}  \\

&\multicolumn{1}{c}{[N\,\textsc{ii}]} &\multicolumn{1}{c}{11.7$\pm$1.9} &\multicolumn{1}{c}{0.4} &\multicolumn{1}{c}{$0.09^{+1.02}_{-0.08}$} &\multicolumn{1}{c}{0.14}  \\

&\multicolumn{1}{c}{[H$\alpha$]} &\multicolumn{1}{c}{18.4$\pm$1.4} &\multicolumn{1}{c}{6.1} &\multicolumn{1}{c}{$0.34^{+1.20}_{-0.33}$} &\multicolumn{1}{c}{0.58} \\

\hline

\multirow{3}{*}{P4} &\multicolumn{1}{c}{[S\,\textsc{ii}]} &\multicolumn{1}{c}{12.7$\pm$2.1} &\multicolumn{1}{c}{1.1} &\multicolumn{1}{c}{$0.22^{+1.61}_{-0.21}$} &\multicolumn{1}{c}{0.22} &\multirow{3}{*}{29.9}\\

&\multicolumn{1}{c}{[N\,\textsc{ii}]} &\multicolumn{1}{c}{13.3$\pm$2.3} &\multicolumn{1}{c}{1.0} &\multicolumn{1}{c}{$0.28^{+0.93}_{-0.27}$} &\multicolumn{1}{c}{0.49} \\

&\multicolumn{1}{c}{[H$\alpha$]} &\multicolumn{1}{c}{19.7$\pm$1.4} &\multicolumn{1}{c}{4.5} &\multicolumn{1}{c}{$0.54^{+1.18}_{-0.52}$} &\multicolumn{1}{c}{0.80}  \\

\hline

\end{tabular*}
\vspace{0.5em}

\begin{minipage}{\textwidth}
\small
$^1$ The Monte Carlo kinematic distance, in units of 100pc\\
$^2$ The method using the rotation curve to compute traditional kinematic distance, in units of 100pc \\ 
\end{minipage}
\end{table*}
%--------------------------------------------------------------------
   \subsection{Electron Temperature and Density}
   \label{subsec:TD}
\begin{figure*}
    \centering
    \includegraphics[width=0.8\textwidth]{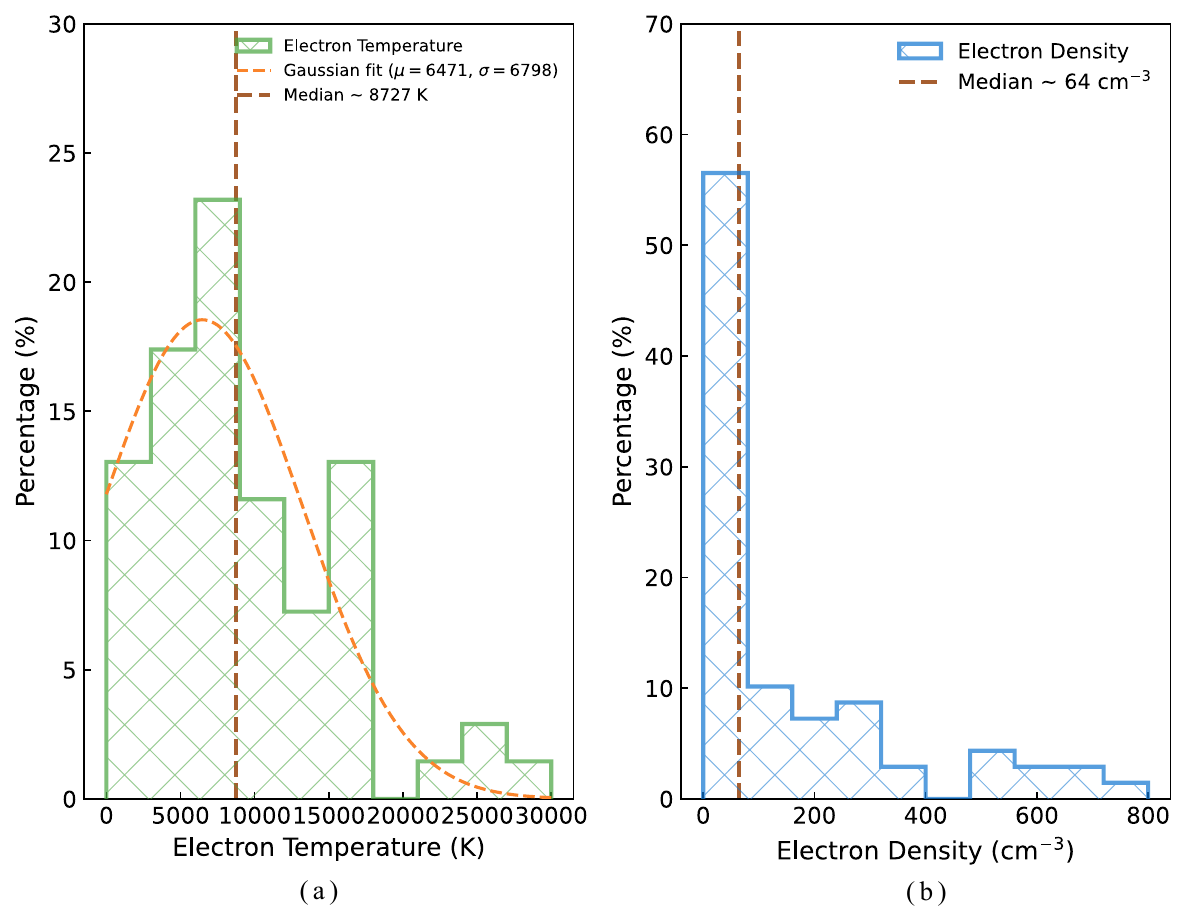}
    \caption{(a) Electron temperatures derived from H$\alpha$ and [N\,\textsc{ii}] line FWHMs. The median Temperature is $\sim 8.7 \times 10^3 \,\mathrm{K}$ (brown dashed line), and the orange dashed curve shows a Gaussian fit with $\sigma \simeq 6.47 \times 10^3 \,\mathrm{K}$, indicating a broad distribution. (b) Electron density ($n_{e}$) in \textit{Silk Nebula} derived from [S\,\textsc{ii}] line ratios. The median density is $\sim 64 \mathrm{cm}^{-3} $ (brown dashed line).}
   \label{fig:nete}
   \end{figure*}

   As the [N\,\textsc{ii}] emission lines are excited by collisions with thermal electrons, the detection of [N\,\textsc{ii}]~$\lambda$6583 indicates that the temperature of the gas is greater than about 3000 K. Once the H$\alpha$ lines are observed simultaneously, the temperature can be limited more restrictively. The FWHMs can provide us more information about the region of interest, such as temperature and velocity. We use the method proposed by \cite{reynolds1977faint} (see Equation (\ref{equ T-FWHM})) to calculate the temperature. 
   \begin{equation}
    \mathrm{T}(\mathrm{K}) = 23.5\mathrm{W}_{\mathrm{H}}^{2}\left(1 - \frac{\mathrm{W}_{\mathrm{N}}^{2}}{\mathrm{W}_{\mathrm{H}}^{2}}\right)\label{equ T-FWHM}
\end{equation}

\begin{equation}
    \mathrm{V}_{\mathrm{t}}(\mathrm{km}\,\mathrm{s}^{-1}) = 1.04\,\mathrm{W}_{\mathrm{N}}\left(1 - 0.071\frac{\mathrm{W}_{\mathrm{H}}^2}{\mathrm{W}_{\mathrm{N}}^2}\right)^{1/2}\label{equ Vt}  
\end{equation}
   where $W_{H}$ and $W_{N}$ are the FWHMs of H$\alpha$ and [N\,\textsc{ii}] emission lines, $T$ is the gas temperature and $V_t$ is the turbulent velocity. Fig~\ref{fig:nete}a presents the frequency distribution of temperature, revealing the median electron temperature is about 9000 K. Such temperatures are common in shock spectra, after rapid cooling, radiative shocks typically reach a quasi-steady-state temperature around $10^4 \,\mathrm{K}$ and maintain an extended recombination cooling zone. The electron temperatures for most pixels covering the [O\,\textsc{iii}] structure fall within the range of 1-2 $\times 10^4~\mathrm{K}$, with a small number of values as high as 2-3 $\times 10^4~\mathrm{K}$. 
   The background region P4 exhibits a median electron temperature of approximately 7.5 $\times 10^3~\mathrm{K}$, which is lower than that of the nebula. In contrast, the eastern background region of the nebula (i.e., RA = 7h49m–7h52m) shows an elevated electron temperature of about $10^4~\mathrm{K}$, exceeding that of the symmetrically located P4 region.
   This may suggest that certain localized regions within the structure have not yet fully radiatively cooled and still retain slightly higher electron temperatures. Overall, we tend to conclude that the average electron temperature of this structure is approximately 10,000 K. $V_{t}$ represents the non-thermal broadening observed in the post-shock region or within cooling shells. When the shock wave enters the ISM with inhomogeneous density, its front end becomes disturbed and torn, inducing structures such as fragments and turbulence in the shell, forming local random velocity fields. These small-scale disturbances generate localized random velocity fields, manifesting as additional line broadening with typical velocities of 10–40 km\,s$^{-1}$. Since their directional and spatial distributions are random, their velocity scales are inherently lower than the shock velocity. As shown in Table~\ref{Table3:V}, the $V_{t}$ in four regions range between 29–33 km\,s$^{-1}$, indicating that the shock-driven gas flows produce localized random velocity fields of tens of km\,s$^{-1}$.

   The [S\,\textsc{ii}] 6716/6732 line ratio is sensitive to electron density and is thus frequently employed as a density diagnostic. We employed the recently updated [S\,\textsc{ii}] electron density diagnostic model from \cite{proxauf2014upgrading} to estimate the electron density (see Equation (\ref{equS/S}) and (\ref{equ ne})), while applying temperature to correct the density following the method described in their work.  
%--------------------------------------------------------------------   
%-------------------------------------------------------------------- 
%--------------------------------------------------------------------   
    \begin{equation}
    \mathrm{R} = \frac{\mathrm{I}_{6716}}{\mathrm{I}_{6732}}\label{equS/S}
\end{equation}

\begin{equation}
\begin{split}
\log(\mathrm{n}_{\mathrm{e}}) &= 0.0543 \tan(-3.0553 \mathrm{R} + 2.8506)\\ 
                 &+ 6.98 - 10.6905 \mathrm{R} + 9.9186 \mathrm{R}^{2} - 3.5442 \mathrm{R}^{3}\label{equ ne}
\end{split}
\end{equation}
   where $I_{6716}$ and $I_{6732}$ are the relative intensities of 6716\AA\, and 6732\AA\,lines respectively, and R is the [S\,\textsc{ii}] line ratio. 
   Meanwhile, we applied corrections for electron temperature in the density calculations to minimize boundary effects. A histogram of the electron density measurements is presented in Fig~\ref{fig:nete}b. For the \textit{Silk Nebula} region, a total of 107 spectral points yield density values peaking between 0 and 100 $\mathrm{cm}^{-3}$. Based on standard atomic physics for [S\,\textsc{ii}] emission under low temperature, low ionization conditions, the mean electron density of the medium is estimated to be $n_{e} \sim 64\,\mathrm{cm}^{-3}$. 
   % \textcolor{red}{
   Notably, all regions of the nebula exhibit enhanced electron densities toward the center, reaching approximately 150 $\mathrm{cm}^{-3}$. In contrast, the [S\,\textsc{ii}] doublet ratio in the outer regions is generally close to the low-density limit ($\sim$ 1.42), indicating electron densities well below 100 $\mathrm{cm}^{-3}$ \citep{osterbrock2006astrophysics}. This spatial trend is broadly consistent with the variation in the [O\,\textsc{iii}] emission-line intensity. By comparison, the background regions show lower electron densities, and the eastern side shows lower densities than those in the nebula itself, and even P4. This is consistent with old supernova remnant shells, whose pre-shocks  typically compress the ambient medium to several times the average density, driving the ahead material via ram pressure.
   % }

   \subsection{Metal abundance}
    The earliest use of emission line intensities as indicators of cosmic elemental abundances can be traced back to \cite{searle1971evidence}, who reported measurements of the nebular [O\,\textsc{iii}] emission in M33. To date, numerous diagnostic indicators have been established, including $N_{2}$, $O_{3}N_{2}$ \citep{pettini2004iii}; $R_{2}$, $O_{32}$, $R_{3}$ \citep{curti2016new}; $N_{2}S_{2}$H$\alpha$ \citep{dopita2016chemical}; and $RS_{32}$, $S_{2}$, $O_{3}S_{2}$ \citep{curti2020mass}. The wavelength coverage of the LAMOST spectra allows us to employ three indirect metallicity calibrators: $N_{2}$, $S_{2}$, and $N_{2}S_{2}$H$\alpha$. The presence of multiple components along the nebular line of sight leads to broadened line profiles and larger uncertainties in the FWHM measurements, which in turn affect the calculation of electron temperature. We therefore exclude the $S_{2}$ index, which is consistently defined on a $T_{\mathrm{e}}$-based abundance scale. For the $N_{2}$ index, we adopt a third-order polynomial of the form (see Equation (\ref{equ N2})) to enable a more precise quantification of the oxygen abundance: 
   \begin{equation}
       12 + \mathrm{log} \mathrm{(O/H)}= 8.9 + 0.57\times\mathrm{N_{2}}\label{equ N2}
   \end{equation}
   where, $\mathrm{N}_{2}$ $\equiv$ log[N\,\textsc{ii}]$\lambda$6583/H$\alpha$. For $\mathrm{N_{2}S_{2}H\alpha}$, the method to estimate the oxygen abundance as follows: 

   \begin{equation}
    \mathrm{y} = \log \frac{[\text{N\,\textsc{ii}}]}{[\text{S\,\textsc{ii}}]} + 0.264 \log \frac{[\text{N\,\textsc{ii}}]}{\text{H}\alpha},  
   \label{equ N2S2}
   \end{equation}   
   \begin{equation}
   12 + \log (\text{O}/\text{H}) = 8.77 + \mathrm{y} + 0.45(\mathrm{y} + 0.3)^5,
   \label{equ O/H}
   \end{equation}

   % \textcolor{purple}{
   The former yields $12 + \log(\mathrm{O/H}) = 8.82$, while the latter gives $8.70$. For N2 index, log[N\,\textsc{ii}]$\lambda$6583/H$\alpha$ exhibits a variable linear relationship with $\log(\mathrm{O/H})$. This calibration is also sensitive to the gas ionization parameter, which introduces systematic uncertainties in the metallicity \citep{zhang2025lamost}. Nonetheless, these values are slightly higher than the solar oxygen abundance of $12 + \log(\mathrm{O/H}) = 8.66$ \citep{prieto2001forbidden}.
   % }
   % \textcolor{red}{
   And using the same index for different regions, including the background, yields relatively consistent results.
   % } 
   The derived oxygen abundance is close to the solar value, which has important implications for determining the nature of the excitation source. First, it indicates no evidence of anomalous nitrogen or oxygen abundances within the structure. This effectively rules out the possibility that the shells were formed by evolved stellar ejecta, such as those from planetary nebulae or Wolf–Rayet ring nebulae, as such shells are typically enriched in helium and nitrogen, exhibit very high [N\,\textsc{ii}]/H$\alpha$ ratios, and deviate from solar metallicities. Combined with the data presented earlier in this paper, these findings support the conclusion that the shock-excited cloud possesses a chemical composition characteristic of the Galactic disk. Moreover, the oxygen abundance measured at different positions shows no significant variation, implying a chemically homogeneous structure. This further supports a shock-driven scenario in which pre-existing ambient gas is heated and ionized by the passing shock, rather than being enriched by freshly synthesized metals.

\section{Discussion}
\label{sect:discu}
\subsection{Is the Silk Nebula really a SNR?}
%-------------------------------------------------------------------- 
\begin{figure*}
\centering
\includegraphics[width=0.9\textwidth]{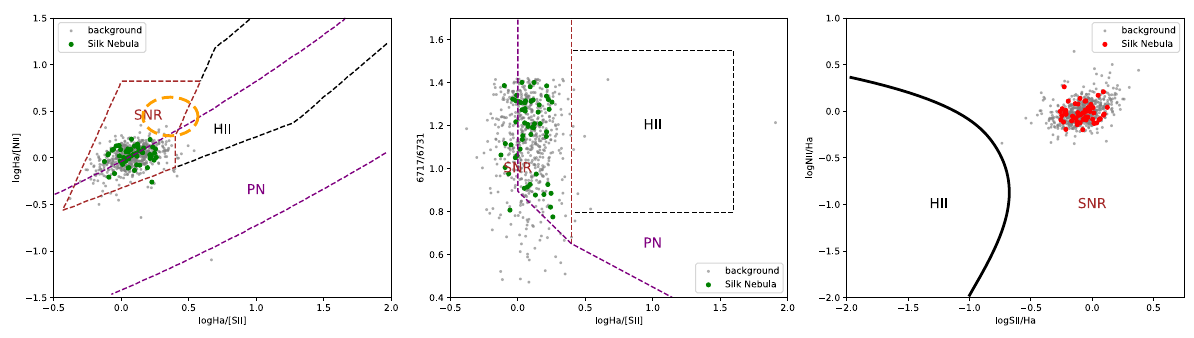}
\caption{Left and middle panels: SNR diagnostic model from \citet{sabin2013new}, where green points mark locations on the [O\,\textsc{iii}] structure and grey points mark the background. The orange circle marks the distribution of DIG regions in the BPT diagrams. Right panel: 2D diagnostic model proposed by \citet{kopsacheili2020diagnostic}, with red points indicating data located on the [O\,\textsc{iii}] structure.}
\label{fig:DIGNOSE}
\end{figure*}
%-------------------------------------------------------------------- 
LAMOST medium-resolution spectra can resolve the H$\alpha$ emission from the adjacent [N\,\textsc{ii}]~$\lambda\lambda$6548,6583 lines, thus avoiding the blending issues common in narrow band imaging and yielding reliable [S\,\textsc{ii}]/H$\alpha$ ratios minimally affected by continuum or filter transmission overlaps. 
We therefore use the optical line ratios to diagnose the nebular type. The ratios of H$\alpha$, [S\,\textsc{ii}], and [N\,\textsc{ii}] are commonly used to distinguish SNRs from H\,\textsc{ii} regions \citep{fesen1985optical}. The most prominent criterion is [S\,\textsc{ii}]/H$\alpha$, with a threshold of $\sim$ 0.45--0.5 typically adopted. According to Table~\ref{Table2:lines}, the [S\,\textsc{ii}]/H$\alpha$ ratio exceeds 0.5 in all four regions, confirming that the ionized gas is shock-excited. The large angular extent and strong [O\,\textsc{iii}]~$\lambda\lambda$4959,5007 emission further rule out photoionization nebulae \citep{skelton1999emission} and giant Herbig--Haro outflows \citep{mader1999new}. However, this is not an exclusive diagnostic: certain compact H\,\textsc{ii} regions and highly excited PNe may also exhibit elevated [S\,\textsc{ii}]/H$\alpha$ ratios \citep{dodorico1980catalogue}. A second indicator is a relatively strong [N\,\textsc{ii}] emission, with [N\,\textsc{ii}]/H$\alpha$ falling in the range 0.5--1.5---a condition satisfied by regions P1, P2, P3, and P4. Meanwhile, a search of SDSS and \textit{Gaia} catalogs reveals no bright or faint central stars near the nebula, while its filamentary morphology and large angular extent effectively rule out a PNe origin.  The [S\,\textsc{ii}]~6716/6732 ratio is sensitive to electron density and older SNRs approach the theoretical limit of $\sim$ 1.42. From the histogram in Fig~\ref{fig:nete}, the electron density of the [O\,\textsc{iii}]-emitting gas peaks in the 0--100~cm$^{-3}$ range, with most values distributed within 0--400~cm$^{-3}$, consistent with the typical density range of SNRs. 
% \textcolor{red}{Such low densities and the absence of radio emission is a characteristic feature of an older supernova remnant.} 

\citet{fesen2024deep} presented deep [O\,\textsc{iii}] and H$\alpha$ imaging of nine known and three newly discovered Galactic SNRs, all but one having at least one angular dimension $>1\degr$. 
Their three new discoveries - G107.7$-$5.1 (the Nereides Nebula, $2.4\degr \times 2.7\degr$ at high Galactic latitude), G209.9$-$8.2 ($1.7\degr \times 1.8\degr$), and G210.5$+$1.3 ($0.4\degr \times 0.3\degr$) --- lie, like the Silk Nebula, toward the Galactic anti-center: G209.9$-$8.2 and G210.5$+$1.3 are at $l \sim 210\degr$, though at Galactic latitudes $b \sim -8\degr$ and $1\degr$ much lower than the Silk Nebula's $b \sim +19\degr$. 
A key finding of their work is that [O\,\textsc{iii}] emission line imaging is essential for mapping the complete shock emission of large, low-surface-brightness SNRs --- precisely the tracer on which our discovery of the Silk Nebula is based. The Silk Nebula fits naturally into this observational class. Its [O\,\textsc{iii}] filament extends over $\sim 10\degr$ on the sky, several times the $1.8\degr$--$2.7\degr$ dimensions of the two $>1.5\degr$ discoveries (G209.9$-$8.2 and the Nereides Nebula), and its [O\,\textsc{iii}] surface brightness of 23.6 -- 23.9~$\mathrm{mag\,arcsec^{-2}}$ (Table~\ref{Table4:Photometry}) corresponds to $\sim 10$--13~Rayleigh --- well above the 0.5 -- 5~Rayleigh detection limit of the \citet{fesen2024deep} [O\,\textsc{iii}] images. So the Silk Nebula has also been detected in Fesen's project \citep{fesen2026optical}, and it is considered to be part of the Monogem Ring. At our fiducial distance of 160~pc, the Silk Nebula's projected extent is $\sim 3~\mathrm{pc} \times 28~\mathrm{pc}$, making it one of the nearest and, in angular extent, one of the largest [O\,\textsc{iii}]-selected filament documented to date.

We further apply the updated SNR diagnostic model of \citet{sabin2013new} and the 2D model of \citet{kopsacheili2020diagnostic} to the \textit{Silk Nebula} data (Fig~\ref{fig:DIGNOSE}). In the left and middle panels, all data points fall securely within the SNR zone, well separated from H\,\textsc{ii} regions and PNe, indicating that the ionization is shock-dominated rather than due to photoionization by young massive stars or evolved stellar wind bubbles.  Spatially, the [O\,\textsc{iii}] emission closely matches the nebular morphology observed in the narrow band images. Although the distributions of [S\,\textsc{ii}], and H$\alpha$ do not show detailed morphological structures apart from the northern extension of P1 and the northern subregion of P2, these data points fall into the SNR in the diagnostic diagrams. Overall, the H$\alpha$ and [S\,\textsc{ii}], emission intensities are significantly stronger on the western side of the nebula than on the eastern side. Hence, on the basis of the analysis presented above, the SNR interpretation remains plausible but is not yet definitive; the most possible conclusion is that Silk is a large-scale, low-surface-brightness [O\,\textsc{iii}] filament, possibly the shell of a nearby evolved supernova remnant. As can be seen from the orange circle in Fig~\ref{fig:DIGNOSE}, the Silk Nebula shows a minor overlap with the DIG region; therefore, the influence of DIG cannot be excluded, even though it is minor (as discussed in detail in Section~\ref{sec4.2}). According to the BPT classification results, the SNR dominates. Based on this conclusion, we estimated the explosion energy and age of the Silk Nebula using SNR models.
%though the possibility of a large-scale DIG structure cannot be ruled out.
% we conclude that the \textit{Silk Nebula} is most likely a supernova remnant.

Based on the fiducial distance of $\sim$ $160\,\mathrm{pc}$ derived in Section~\ref{subsec:distance} and the curvature of the nebula, the physical radius($R_{\text{s}}$) of the complete remnant is estimated to be approximately $30\,\mathrm{pc}$. The post-shock shell electron density is $n_e \sim 64$--$150\,\mathrm{cm}^{-3}$ from the [S\,\textsc{ii}] line ratio (Section~\ref{subsec:TD}). For an evolved, low-density SNR shell, the pre-shock ambient density is expected to be substantially lower; we adopt a fiducial value of $n_0 \sim 1.0\pm0.3\,\mathrm{cm}^{-3}$. Following the relation between preshock ambient density and current shock’s velocity proposed by \citet{dopita1979optical}, the current shock velocity $\sim$ 119$~\pm~5$~km~s$^{-1}$.
Its large spatial scale and spectral features indicate that the \textit{Silk Nebula} has evolved beyond the Sedov-Taylor phase and entered the pressure-driven snowplow (PDS) phase. To verify this hypothesis, we estimated the critical radius $R_{\text{PDS}}$ for the transition from the Sedov-Taylor to the PDS phase \citep{cioffi1988dynamics}: $R_{\text{PDS}} =14.0 ( \frac{E}{10^{51} \text{ erg}} )^{\frac{2}{7}} ( \frac{n_0}{\text{cm}^{-3}} )^{-\frac{3}{7}} \xi_m^{-\frac{1}{7}}$
where $\xi_m$ is a constant equal to unity for solar metallicity. In the PDS phase, following \citet{boumis2022deep} the energy is found to be $E = 6.8 \times 10^{43} (\frac{n_0}{\text{cm}^{-3}})^{1.16} (\frac{V_s}{\text{km s}^{-1}})^{1.35} (\frac{R_s}{\text{pc}})^{3.16} \xi_m^{-0.161} = (2.0 \pm 0.7) \times 10^{51} \text{ erg}$, which yields a critical radius of 18 $\pm~2.8$ pc. This value is smaller than the current radius of the nebula, indicating that the remnant has indeed entered the PDS phase, leading to an estimated age $t = 9.98 \times 10^3 \times (\frac{E}{10^{51} \text{ erg}})^{\frac{3}{14}} (\frac{n_0}{\text{cm}^{-3}})^{-\frac{4}{7}} \xi_m^{-\frac{5}{14}} [(\frac{R_s}{R_{\text{PDS}}})^{\frac{10}{3}} + 1]$ = $(6.8\pm1.8)\times 10^4\,\mathrm{yr}$. 
We searched the NRAO VLA Sky Survey (NVSS; \cite{condon1998nrao}) and the TIFR GMRT Sky Survey (TGSS; \cite{intema2017gmrt}) and found no radio counterpart. The absence of radio emission is consistent with the old age ($\sim 6.8 \times 10^4$ yr) estimated for this remnant, as evolved SNRs typically have low radio surface brightness. %However, a non-SNR origin cannot be excluded on this basis alone.
% We adopt a supernova explosion energy of $E \sim 2.25 \times 10^{51}\,\mathrm{erg}$ following \citet{boumis2022deep}, and the critical radius is 18.3 pc\citep{boumis2022deep}, smaller than the current radius. It confirms that the remnant has indeed entered the PDS phase, leading to an estimated age of $\sim 6.8\times 10^4\,\mathrm{yr}$. 

\subsection{DIG contamination}\label{sec4.2}
However, there remains a question: within the same region of the nebula, we detected a systematic velocity offset of 5--15~km~s$^{-1}$ between the [O\,\textsc{iii}] emission and the [N\,\textsc{ii}], [S\,\textsc{ii}], and H$\alpha$ lines, corresponding to $V_{\rm LSR}$ differences of 10--16~km~s$^{-1}$. This leads to a distance discrepancy of 100--150~pc between kinematic distance estimates from the blue and red arm lines, whereas the radial velocities and derived distances for individual emission lines remain highly consistent across different subregions. The turbulent velocity component ($V_{\rm t}$), derived primarily from the [N\,\textsc{ii}] linewidth after thermal broadening correction (because heavier nitrogen lines are more sensitive to non-thermal motions than hydrogen), ranges from 29 to 33~km~s$^{-1}$ across the four subregions (Table~\ref{Table3:V}), demonstrating good internal agreement. Given these discrepancies, we cannot rule out the possibility of another ISM component superimposed along the line of sight to the nebula. Based on the high [S\,\textsc{ii}]/H$\alpha$ and [N\,\textsc{ii}]/H$\alpha$ ratios observed in the spectra, we conclude that this additional component is most plausibly DIG.

The uncertainty in classifying this nebula as a SNR arises in part from DIG contamination in the red arm. This interpretation is supported by three lines of evidence: (1) the RVs of the red arm emission lines (H$\alpha$, [N\,\textsc{ii}], and [S\,\textsc{ii}]) toward the nebula are indistinguishable from those of the background region; (2) the [S\,\textsc{ii}]/H$\alpha$ ratios are comparably high in the background; and (3) the [S\,\textsc{ii}] and H$\alpha$ emissions lack discernible morphological structures in the narrow-band images. 
To assess the contribution of DIG along the line of sight, we apply the diagnostic criteria of \cite{wen2025diffuse} for distinguishing DIG in the Galactic anti-center direction. As listed in Table~\ref{Table2:lines}, the background region P4 shows [N\,\textsc{ii}]/H$\alpha > 0.7$ and [S\,\textsc{ii}]/H$\alpha > 0.8$, far exceeding the DIG threshold of 0.24 given in \cite{wen2025diffuse} and consistent with the low-ionization character of DIG. The velocity differences among H$\alpha$, [N\,\textsc{ii}], and [S\,\textsc{ii}] in this region remain below 20~km~s$^{-1}$, and the Galactic longitude falls within the applicable range of the diagnostic. 

For completeness, we also examined the typical DIG locus in the BPT diagram using the distribution from \citet{belfiore2022tale} (shown as the orange region in Fig.~\ref{fig:DIGNOSE}). This DIG locus is concentrated near the demarcation line between the SNR and H\,\textsc{ii} regimes, and a modest fraction of our data points lie within or near this region. We therefore infer that the DIG contributes to the observed line ratios, particularly in the red arm, and constitutes a systematic uncertainty in both the kinematic distance estimates and the SNR classification.

We note that even if DIG contamination were absent, the $\sim$5--15 km s$^{-1}$ velocity offset between [O\,\textsc{iii}] and the low-ionization lines would still require explanation. One possibility is that the [O\,\textsc{iii}] and H$\alpha$ emissions arise from different layers of the same expanding shell---for example, a faster outer shock front and a decelerated inner cooling region. The observed offset can be naturally accommodated by projection effects, as both layers share a similar large-scale velocity field and their differential motion projected onto the line of sight need only be a small fraction of the total expansion speed ($\sim$140 km s$^{-1}$, Section~\ref{subsec:TD}). Alternatively, the nebula may be an inhomogeneously expanding bubble in which [O\,\textsc{iii}] and H$\alpha$ originate from physically distinct regions whose local expansion velocities differ, producing the $\sim$10--15 km s$^{-1}$ offset without requiring a large intrinsic velocity contrast. Both scenarios are physically plausible but cannot be distinguished with the current data, and neither excludes a contribution from DIG. The combined uncertainties from DIG contamination and the unresolved velocity structure underscore why the SNR classification remains provisional at this stage.

\section{Conclusion}
\label{sect:concl}
% \textcolor{red}{
The \textit{Silk Nebula} was discovered and named by the SWONS team. The first complete multi-fiber, medium-resolution spectroscopy of the nebula was subsequently obtained through the LAMOST MRS-N project, allowing us to derive its key physical parameters. 
The shock diagnostic ratio [S\,\textsc{ii}]/H$\alpha$ significantly exceeds the classical threshold of 0.4. In the BPT diagnostic diagrams, the data points lie within the SNR regime, though a modest fraction falls near the DIG locus, and the elevated [S\,\textsc{ii}]/H$\alpha$ ratios in the background region indicate that DIG contamination cannot be entirely ruled out.
% the data points lie entirely within the SNR regime and show negligible overlap with the DIG region. 
Moreover, the spatial distribution of the [O\,\textsc{iii}] emission closely follows the narrowband imaging morphology, consistent with the characteristics of shock-excited SNR. We suggest that the Silk Nebula is best described as a large [O\,\textsc{iii}] filament of possible supernova-remnant origin, powered by shock-heated gas; a DIG or ionized-bubble interpretation cannot be fully excluded with the current data.
% We therefore infer that the large-scale \textit{Silk Nebula} is a part of the SNR powered by shock-heated gas. 
Elemental abundances derived from the $N_{2}$ and $N_{2}S_{2}$H$\alpha$ indices are near-solar and spatially uniform, further supporting a scenario in which a shock wave propagates into the ambient medium, heating and ionizing the pre-existing gas. 
% The Silk Nebula was discovered and named for the first time in the SWONS team. Subsequently, the first complete multifiber,  medium-resolution spectroscopy of the nebula were secured through the LAMOST MRS-N project, yielding its key physical parameters. 
% The shock diagnostic ratio [S\,\textsc{ii}]/H$\alpha$ is significantly greater than the classical threshold of 0.4. In the BPT diagnostic diagrams, the data points lie entirely within the SNR regime, showing negligible overlap with the DIG region. Furthermore, the spatial distribution of the [O\,\textsc{iii}] emission line aligns well with the narrowband imaging morphology, consistent with shock-excited SNR features. Consequently, we infer that the optical emission originates from shock-heated gas. The elemental abundances derived from the $N_{2}$ and $N_{2}S_{2}$H$\alpha$ indices are near-solar and exhibit a uniform spatial distribution, which further supports the scenario where a shock wave propagates into the ambient medium, heating and ionizing the surrounding pre-existing gas.
% }

% \textcolor{red}{
The line-of-sight spectral data exhibit prominent emission lines of H$\alpha$, [N\,\textsc{ii}] $\lambda\lambda$6548,6583, and [S\,\textsc{ii}] $\lambda\lambda$6716,6732, alongside moderate [O\,\textsc{iii}] $\lambda\lambda$4959,5007 emissions. In this study, we selected and analyzed 2,684 high-quality fiber spectra covering the \textit{Silk Nebula} (including its peripheral regions). By measuring parameters such as relative line intensity, $V_{\rm LSR}$, and FWHM, we established a comprehensive spectral dataset for the \textit{Silk Nebula}. 
% }

% \textcolor{red}{
From the $V_{\rm LSR}$ and based on the [O\,\textsc{iii}] kinematics alone, we estimate a fiducial distance of $\sim$160 pc using both Monte Carlo and traditional kinematic distance methods. At this distance, the angular extent of $\sim 10^\circ$ corresponds to a physical scale of $\sim$28 pc along the north-south direction, making it potentially one of the largest nebulae in terms of angular extent documented to date.
However, owing to the velocity discrepancies among different emission lines and the proximity of the nebula, these estimations are subject to certain systematic uncertainties.
% }

% \textcolor{red}{
The electron density derived from the [S\,\textsc{ii}] doublet ratio ranges from several tens to $\sim 100\ \mathrm{cm}^{-3}$, consistent with the characteristic shell density of SNRs. Adopting a fiducial pre-shock density of $n_0 \sim 1.0\pm0.3\,\mathrm{cm}^{-3}$ for an evolved SNR shell, which satisfies the low-density condition required for shell expansion. The shock velocity is estimated to be $119\pm5\ \mathrm{km\ s}^{-1}$. 
The absence of detectable radio counterparts in NVSS and TGSS is consistent with the interpretation of the nebula as an evolved SNR in the PDS phase, for which low radio surface brightness is expected at ages of $\sim 10^4$--$10^5$ yr. 
Under this assumption, we derive a shell energy of $(2.0\pm0.7) \times 10^{51}\ \mathrm{erg}$ and a current age of $(6.8\pm1.8) \times 10^4\ \mathrm{yr}$.
While our study tentatively identifies this large-scale [O\,\textsc{iii}] structure as a shock-excited SNR shell, further investigations are required to firmly establish its nature. 
% }

% \textcolor{red}{
Here, we outline future observational and modeling efforts aimed at placing tighter constraints on its physical properties. 
H$\beta$ $\lambda$4861 \AA\ lies outside the 4950–5350 \AA\ wavelength coverage of the LAMOST MRS-N blue arm, so the nebular [O\,\textsc{iii}]/H$\beta$ ratio is not accessible with the present data. Future spectroscopy extending blueward of 4950 \AA \ to cover H$\beta$ would allow the [O\,\textsc{iii}]/H$\beta$ ratio to be measured directly and place the shell on the BPT diagram, enabling an assessment of whether the shell emission is photoionized or shock-excited. A high ratio ([O\,\textsc{iii}]/H$\beta \gg$ 1), in the absence of other hard photoionizing sources, would lend further support to a shock or precursor.
In the X-ray regime, middle-aged supernova remnants often exhibit faint, diffuse soft X-ray emission arising from hot, tenuous interior plasma. Deep exposures of this field with current sensitive X-ray observatories could search for this component; the detection of extended soft X-ray emission spatially coincident with the optical shell would provide compelling evidence for a supernova remnant origin, while also constraining the temperature and density of the hot gas. Additionally, quantitatively modeling the observed emission-line ratios with state-of-the-art shock codes — by varying key parameters such as shock velocity, pre-shock density, magnetic field strength, and pre-ionization conditions — will yield deeper insight into the physical conditions and environment of this candidate remnant.
% Here, we outline prospective observational and modeling efforts to place tighter constraints on its physical properties. In the X-ray regime, middle-aged SNRs often exhibit faint, diffuse soft X-ray emission originating from hot, uncooled interior plasma. Deep exposures of this field with current sensitive X-ray observatories could search for this component; detecting extended soft X-ray emission spatially coincident with the optical shell would provide compelling evidence for its SNR origin, while also constraining the temperature and density of the hot gas. Additionally, quantitatively modeling the observed emission line ratios with state-of-the-art shock codes - by varying key parameters such as shock velocity, pre-shock density, magnetic field strength, and pre-ionization conditions---will yield a deeper understanding of the physical conditions and the nebular environment of this candidate remnant.
% }
\begin{acknowledgements}
~This work is supported by the National Key R\&D Program
of China (2021YFA1600401 and 2021YFA1600400), the
National Natural Science Foundation of China (NSFC; grant
Nos. 12090041, 12090040, 12273052, 11733006, U1931109
and 12003043), and the science research grants from the China
Manned Space Project (Nos. CMS-CSST-2025-A14, CMS-CSST-2025-A19 and
CMS-CSST-2021-A04).

J.Z. acknowledges NSFC grant 12425302, and support by Hong Kong Jockey Club Charities Trust through JC STEM Lab of Astronomical Instrumentation.

Z-H.H acknowledges NSFC grant 12573023,12303024. 

This work made use of the data from LAMOST (Large Sky Area Multi-Object Fiber Spectroscopic Telescope, also known as the Guoshoujing Telescope) (\url{https://cstr.cn/31118.02.LAMOST}). LAMOST is a Chinese national mega-science facility, operated by National Astronomical Observatories, Chinese Academy of Science.
\end{acknowledgements}

\bibliography{sample701}{}
\bibliographystyle{aasjournalv7}



\end{document}